\documentclass[12pt]{article}
\usepackage{amsmath}
\usepackage{amssymb}
\usepackage{bm}
\usepackage{graphicx}
\usepackage{enumerate}
\usepackage{natbib}
\usepackage{url} 
\usepackage{caption}
\usepackage{subcaption}
\usepackage[ruled,vlined]{algorithm2e}

\def\eb{\overline{E}}
\newcommand{\blind}{1}
\usepackage[dvipsnames]{xcolor}
\usepackage{tikz}
\newcommand{\tikzcircle}[2][red,fill=red]{\tikz[baseline=-0.5ex]\draw[#1,radius=#2] (0,0) circle ;}

\begin{document}

\def\spacingset#1{\renewcommand{\baselinestretch}%
{#1}\small\normalsize} \spacingset{1}


\if1\blind
{
  \title{\bf Bayesian Structural Learning with Parametric Marginals for Count Data: An Application to Microbiota Systems}
  \author{Veronica Vinciotti\thanks{These authors contributed equally to this work.}\hspace{.2cm}\\
    Department of Mathematics, University of Trento\\
    and \\
    Pariya Behrouzi$^*$\\
    Department of Mathematics, Wageningen University \\
		and \\
		Reza Mohammadi\\
    Faculty of Economics and Business, University of Amsterdam
		}
		\date{}
  \maketitle
} \fi

\if0\blind
{
  \bigskip
  \bigskip
  \bigskip
  \begin{center}
    {\LARGE\bf Bayesian Structural Learning with Parametric Marginals for Count Data: An Application to Microbiota Systems}
\end{center}
  \medskip
} \fi

\begin{abstract}
High dimensional and heterogeneous count data are collected in various applied fields. In this paper, we look closely at high-resolution sequencing data on the microbiome, which have enabled researchers to study the genomes of entire microbial communities. Revealing the underlying interactions between these communities is of vital importance to learn how microbes influence human health. 
To perform structural learning from multivariate count data such as these, we develop a novel Gaussian copula graphical model with two key elements. Firstly, we employ parametric regression to characterize the marginal distributions. This step is crucial for accommodating the impact of external covariates. Neglecting this adjustment could potentially introduce distortions in the inference of the underlying network of dependences. 
Secondly, we advance a Bayesian structure learning framework, based on a computationally efficient search algorithm that is suited to high dimensionality. The approach returns simultaneous inference of the marginal effects and of the dependence structure, including graph uncertainty estimates. A simulation study and a real data analysis of microbiome data highlight the applicability of the proposed approach at inferring networks from multivariate count data in general, and its relevance to microbiome analyses in particular. The proposed method is implemented in the R package \texttt{BDgraph}.
\end{abstract}

\section{Introduction}
\label{sec:intro}
Graphical modelling approaches allow to learn statistical dependences from multivariate data. Among these, Gaussian graphical models are by far the most popular, thanks also to their efficient implementations for high dimensional problems \citep{friedman08,mohammadi21}. In many applied fields, however, data are far from Gaussian. In this paper, we consider the case of count data, such as the  high-resolution sequencing data collected routinely in genomic studies. It is not uncommon for these data to feature marginal distributions that are skewed and with a large mass at zero. For this reason, transformations, such as the logarithm or the centered log ratio, are typically applied to genomic data, followed by Gaussian graphical modelling approaches on the transformed data. This is for example the case of the two most used methods for microbiome data, SparCC \citep{friedman12} and SPIEC-EASI \citep{kurtz15}. 
These transformations require a pseudo-count adjustment to be able to handle zeros and may therefore impact also the network inference conducted downstream. 

In the literature, extensions of Gaussian graphical models to non-Gaussian data can take different forms but there is generally little research for the case of unbounded count data, such as the genomic data that we discuss above. \cite{Roy20} have recently proposed a pairwise Markov random field model with flexible node potentials, while \cite{cougoul19} have proposed a Gaussian copula graphical model to couple the contribution from the marginal distributions with that of the underlying dependence structure. Considering microbiota systems as the specific application, they propose zero-inflated negative Binomial marginals.  Our work is linked to this second paper. On the one hand the use of a Gaussian copula facilitates the integration of novel approaches with existing ones that rely on Gaussianity, without the need for ad-hoc transformations. On the other hand, the use of parametric marginal distributions, rather than the non-parametric empirical distributions as in the popular non-paranormal approach \citep{liu09}, 
facilitates the inclusion in the model of  additional covariates, which are, for example, typically available in genomic studies but often ignored due to methodological restrictions. These marginal effects, if left unaccounted for, could distort the inference of the underlying network of dependences.

While there are no specific constraints in the choice of the parametric marginal model, in this paper, we advocate the use of discrete Weibull regression for linking the marginal distributions to external covariates \citep{klakattawi18,haselimashhadi18,peluso19}. The simplicity of this distribution (a two-parameter distribution), combined with the fact that the two parameters can jointly capture broad levels of dispersion (from under to over), makes it quite an appealing candidate for  multivariate count data with a high number of random variables and/or external covariates, such as the microbiota data that we consider in the real data example. This is because, firstly, for a large number of count variables, one wants to avoid tuning the type of distribution for each variable, and, secondly, for a large number of external covariates, a global requirement of over dispersion at all levels of the covariates could
prove too restrictive. Finally, an important feature of the discrete Weibull distribution in the context of Gaussian copula graphical models, is the fact that it is generated as a discretized form of a continuous Weibull distribution (see Figure \ref{fig:dwcwlink}). This creates a latent non-Gaussian space in the vicinity of the data, with a one-to-one mapping with the latent Gaussian data, where the conditional independence graph resides. 

A fundamental problem of copula graphical models for discrete data, bounded or unbounded, is the fact that the marginal distributions are not strictly monotonic. In this setting, while the existence of a copula can still be guaranteed by Sklar's theorem \citep{sklar59}, its uniqueness can not. In fact, the class of copulas compatible with a given discrete dataset can be quite large, leading to potential biases in the inferential procedure \citep{genest07}. On the one hand, this problem is alleviated by the presence of covariate dependent marginals, particularly when the covariates are continuous and the underlying network does not depend on the covariates \citep{yang20}. This can be seen as a second advantage of incorporating covariates in the marginal models, when the objective is to perform structural learning for count data. 
On the other hand, more advanced inferential procedures are required, that account for the fact that each observed count is associated with an interval in the latent Gaussian space. This relies on the ideas of extended rank likelihood \citep{hoff07} and has been used also in the context of Gaussian copula graphical models, both in a frequentist setting \citep{behrouzi19} and in a Bayesian setting \citep{dobra11,dobra18,mohammadi17,murray13}.  While extended rank likelihood has been developed for ordinal (bounded) data, in this paper we develop these approaches for Gaussian copula graphical models on unbounded count data with parametric marginals. The use of these approaches avoids the need for ad-hoc data transformation procedures that condense each interval into one point, with choices such as the right-most point of the interval (essentially using the non-paranormal approach of \cite{liu09,liu12} on count data) or the point corresponding to the median of the distribution function at the two extremes of the interval \citep{cougoul19}. These choices, while efficient, may not work well with skewed distributions, or generally distribution functions that are highly stepwise. 

Finally, we conduct inference in a Bayesian framework, leading to a novel Bayesian structure learning procedure in the context of Gaussian copula graphical models with parametric marginals, extending the efficient computational approaches that have been recently proposed for this class of models \citep{mohammadi15,mohammadi21}, and providing an alternative to frequentist approaches \citep{cougoul19}. Appropriate choices of a prior distribution on the graphs can be made to encourage sparsity. Importantly, uncertainty on the graph learning is fully quantified by the procedure and can be summarized in various ways, such as by calculating posterior probabilities for each edge via Bayesian averaging. This  plays a crucial role, particularly in high dimensional settings, where model selection methods for regularized approaches do not work well and where there is typically a large uncertainty around the optimal graph.

In conclusion, this paper presents a novel methodology for structural learning from high dimensional heterogeneous count data. 
 Section \ref{sec:method} will describe the details of the methodology proposed, whose implementation has been included in the R package {\tt BDgraph} \citep{mohammadi19}. A simulation study in Section \ref{sec:simulation} and a real data analysis of microbiome data from the Human Microbiome Project \citep{HMP} in Section \ref{sec:realdata} will show the usefulness of the proposed approach at inferring networks from high dimensional count data in general, and in the context of microbiota data analyses in particular. Finally, Section \ref{sec:conclusion} will draw some conclusions and point to future research directions.

\section{Methods}
\label{sec:method}
In this section, we present the technical details of the proposed method, starting with the definition of a Gaussian copula graphical model and of the discrete Weibull (DW) regression used for the marginal components, followed by the Bayesian inferential procedure.
\subsection{Gaussian copula graphical model with DW marginals}
Let $\bm{Y} = (Y_{1},\ldots, Y_{p})$ be a vector of count variables. In the case of microbiota systems that we consider in Section \ref{sec:realdata}, these are abundances of the individual microbes or, more commonly, of the Operating Taxonomic Units (OTUs) into which they are clustered, e.g., bacterial species. Let $F_j(\cdot)$, $j=1,\ldots,p$, be the cumulative distribution functions associated to the $p$ variables, respectively.  In a copula graphical model, the joint distribution of the variables is described via a copula function $C(\cdot)$ that couples the marginal distributions $F_j(\cdot)$ into their joint dependence. Formally, 
\begin{equation*}
\label{eq:copula}
P(Y_{1} \leq y_{1},\ldots, Y_{p} \leq y_{p}) = C(F_{1}(y_{1}),\ldots,F_{p}(y_{p})~| \boldsymbol{\Theta}),
\end{equation*}
where $\boldsymbol{\Theta}$ are the parameters describing the copula function $C(\cdot)$. In the case of a Gaussian copula	\citep{hoff07,mohammadi17}	
\begin{equation*}
\label{eq:Gaussiancopula}
P(Y_{1} \leq y_{1},\ldots, Y_{p} \leq y_{p}) = \Phi_{p}(\Phi^{-1}(F_{1}(y_1)), \ldots, \Phi^{-1}(F_{p}(y_p))| \bm{R}),
\end{equation*}
where $\Phi_{p}(\cdot)$ is the cumulative distribution function of a $p$-dimensional multivariate normal with a zero mean vector and  correlation matrix $\bm{R}$, while $\Phi(\cdot)$ is the standard univariate normal distribution function. The dependence structure is captured by the inverse of the correlation matrix $\bm{K} = \bm{R}^{-1}$, typically called the precision or concentration matrix. In particular, 
the zero patterns in this matrix define the conditional independence graph in the latent Gaussian space, following from the theory of Gaussian graphical models \citep{lauritzen96}.

In the context of copula graphical models, the marginal distributions $F_j(\cdot)$ are typically considered as nuisance parameters and estimated by their empirical counterpart. However, in real-world applications, such as in genomic studies, external covariates are often available and there is an interest in estimating their effect on the outcome while accounting for the multivariate nature of the data. In this paper, we argue how, accounting for external covariates at the marginal level is important also for structural learning. Indeed, when the dependence structure does not vary with the covariates, adjusting for marginal effects has the two benefits of widening the range of the marginal distributions at each discrete point and of correcting for the bias in the estimation of multivariate dependences induced by the marginal effects, respectively. 

In this paper, we propose to model the marginal components, and their link with covariates, via a discrete Weibull regression \citep{peluso19}, although other count distributions can be used at this stage. Formally, let $\bm{X}=(1,X_1,\ldots,X_d)^t$ be a vector of covariates. Then, the conditional distribution of $Y_j$ given $\bm{X}$ is modelled by:
\begin{equation}
\label{eq:dwmarginals}
F_{j}(y_j|\bm{X}=\bm{x})=1-q_j(\bm{x})^{(y_j+1)^{\beta_j(\bm{x})}}, \quad y_j=0,1,\ldots,
\end{equation}
where the function $q_j(\cdot)$, corresponding to the parameter $q$ of the distribution, takes values between 0 and 1, while $\beta_j(\cdot)$ is associated to the parameter $\beta$ and takes values in the positive real line. We link the parameters to the external covariates using the logit and the log links, respectively, that is
\begin{equation}
	\label{dw_linkfun}
 q_j(\bm{x}) = \dfrac{\exp(\bm{x}^t{\bm \theta_j})}{1+\exp(\bm{x}^t{\bm \theta_j})},
 \quad \beta_j(\bm{x}) = \exp(\bm{x}^t{\bm \gamma_j}),
\end{equation}
with ${\bm \theta_j}$ and ${\bm \gamma_j}$ denoting the regression coefficients associated to the $Y_j$ marginal component of the model. For other choices of link functions, see \cite{haselimashhadi18}. The simplest case of only the intercept in each model corresponds to the case of no external covariates, i.e., simply discrete Weibull marginal distributions. 
In the real data analysis, we also consider a model with an additional zero inflation component $\pi_j(\bm{x})$. This is common in the microbiome literature due to the sparsity of the data \citep{cougoul19}, although we find that this zero-inflated model is rarely selected against the simpler model.

\begin{figure}[!h]
        \begin{center}
            \begin{subfigure}{0.32\textwidth}
               \includegraphics[scale=0.3]{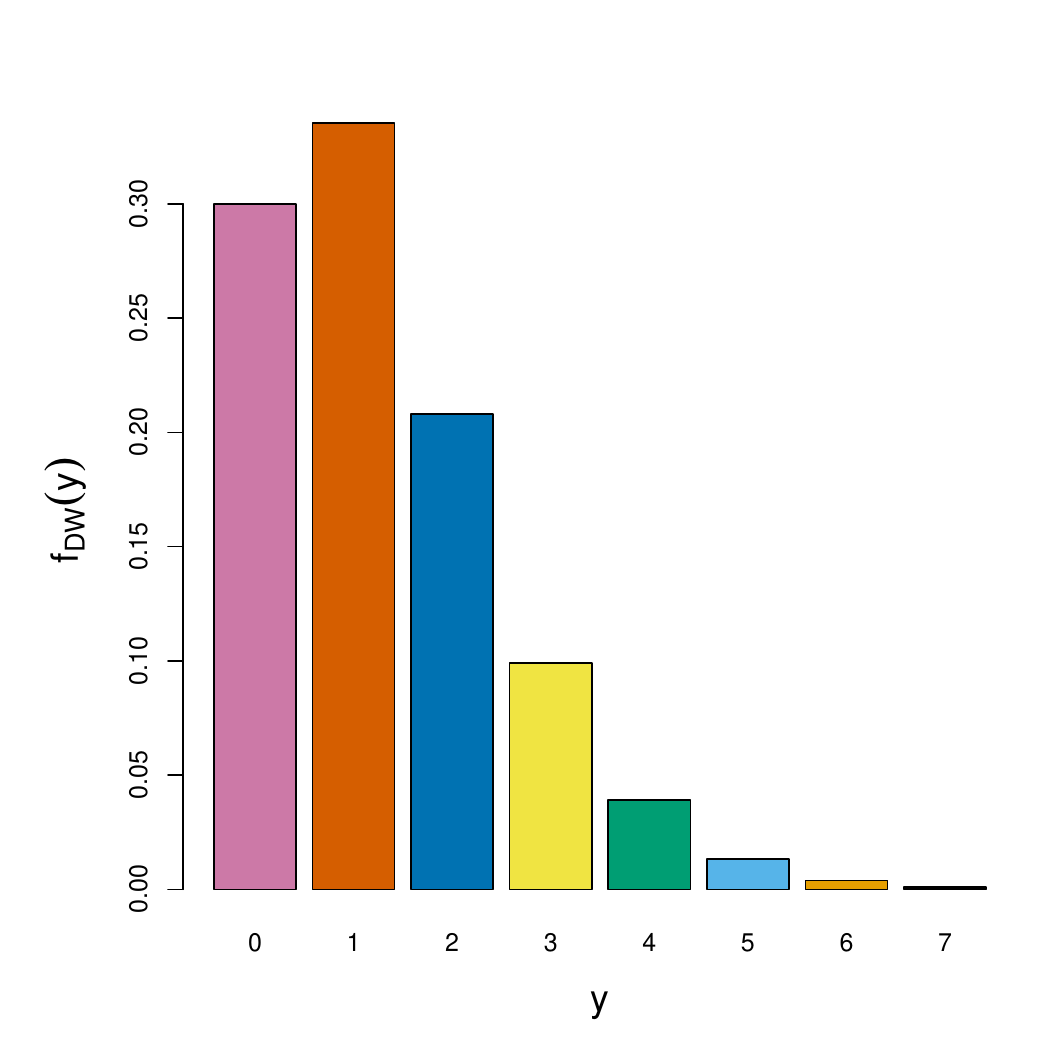} 
                \subcaption{Discrete Weibull}
                \label{fig:dwcwlink1}
            \end{subfigure}
            \hfill
            \begin{subfigure}{0.32\textwidth}
            \centering
               \includegraphics[scale=0.3]{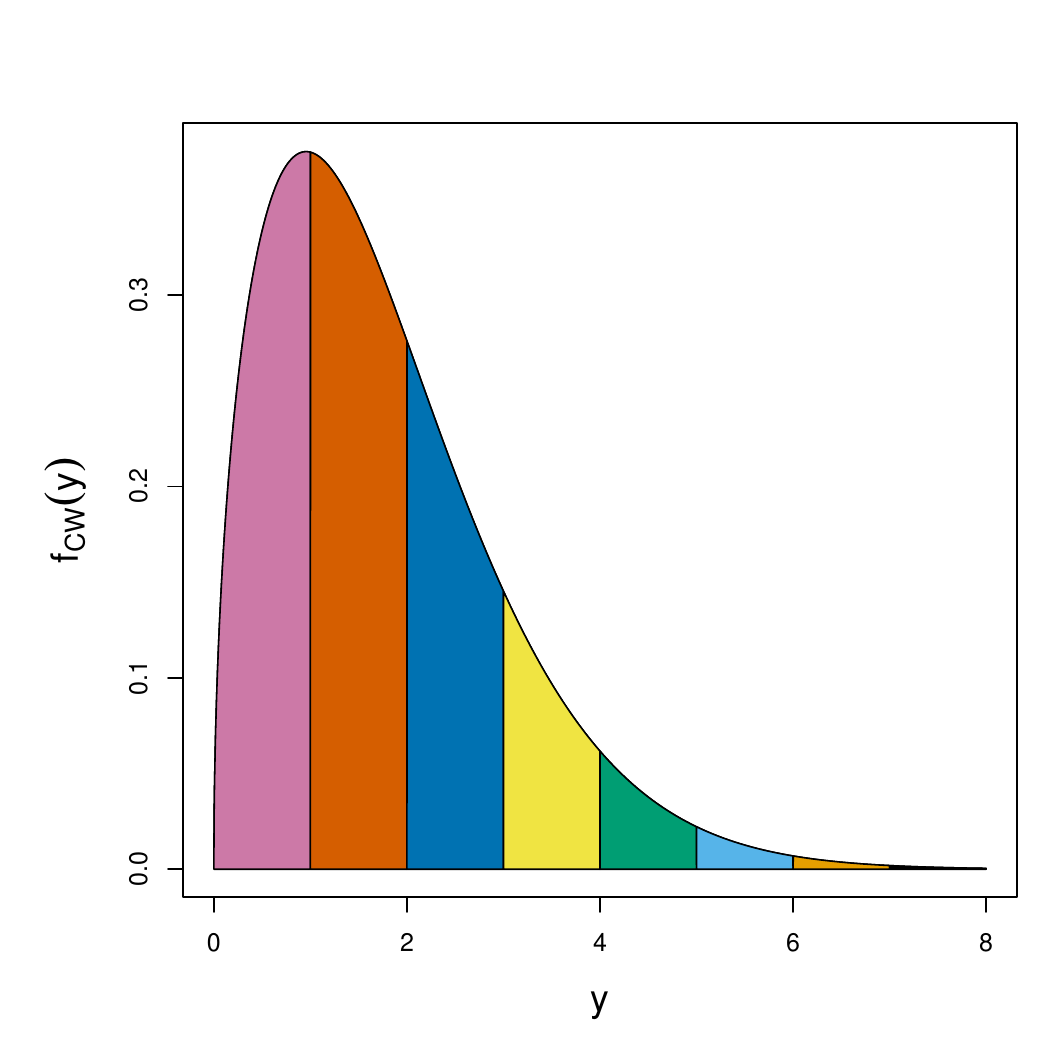} 
                \subcaption{Latent Continuous Weibull}
                \label{fig:dwcwlink2}
            \end{subfigure}
						\hfill
            \begin{subfigure}{0.32\textwidth}
            \centering                
						\includegraphics[scale=0.3]{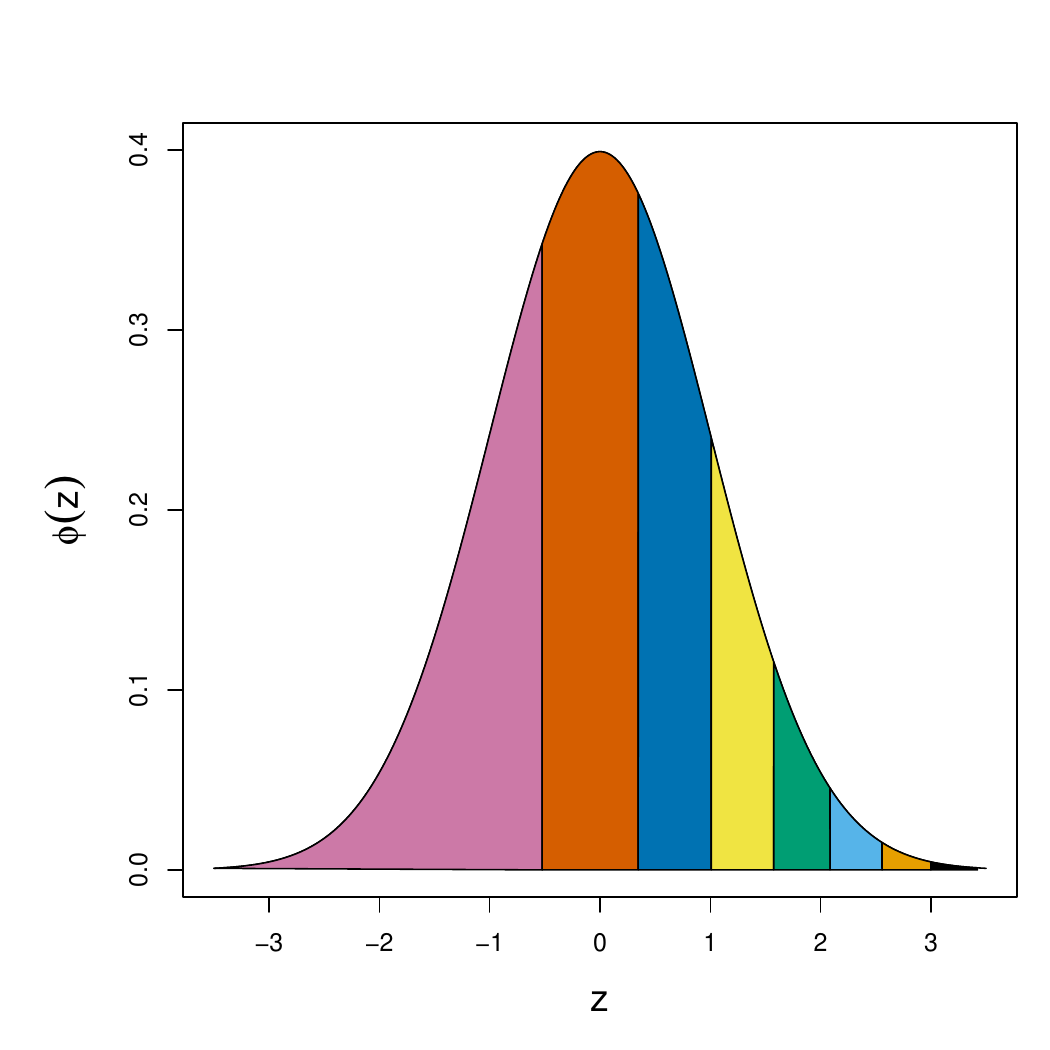}
						 \subcaption{Latent Gaussian}
                \label{fig:dwcwlink3}
            \end{subfigure}
        \end{center}
 \caption{Schematic connection between (a) the discrete Weibull probability mass function $f(y;q=0.7,\beta=1.5)$, (b) the underlying continuous Weibull density $f_{CW}(y; q,\beta)$ and (c) the latent Gaussian $z=\Phi^{-1}(F_{CW}(y; q,\beta))$. Each colour relates to the probability associated to the corresponding value (here microbial abundance) of the discrete random variable.}   
\label{fig:dwcwlink}
\end{figure}

A few properties of a discrete Weibull distribution make it an ideal candidate for modelling high dimensional count data. In particular:
\begin{enumerate}
\item $F_{j}(0|\bm{X}=\bm{x})=P(Y_j=0|\bm{X}=\bm{x})=1-q_j(\bm{x})$, thus the parameter $q$ models directly the proportion of zeros in the data and the effect of covariates on this. This may be useful for datasets with a large percentage of zeros. 
\item The two parameters of the distribution are sufficient to capture both under and over dispersion levels, while still being a parsimonious choice (e.g., same number of parameters as the commonly used negative Binomial distribution). This has been shown to be useful on real data analyses of count data, particularly in the case of under dispersion \citep{peluso19}. Although micriobiome data are typically highly over dispersed, the presence of a large number of external covariates could make a global requirement of over dispersion at all levels of the covariates $\bm{x}$ too restrictive. Moreover, capturing both over and under dispersion is appealing when modelling any generic high dimensional multivariate count data, as it avoids the fine tuning of the most appropriate marginal distribution for each variable.  
\item The quantiles of the distribution have a closed-form expression, with the $\tau$ quantile, for $\tau \ge 1-q$, given by \cite{peluso19},
\[\mu_{(\tau)}=\Big\lceil \biggl(\dfrac{\log(1-\tau)}{\log(q)} \biggr)^{1/{\beta}}-1 \Big\rceil,
\]
where $\lceil \cdot \rceil$ denotes the ceiling function. This means that a re-parametrization based on the median is also possible, e.g., when quantification of the covariate effects is of primary interest \citep{burger20}.
\item The distribution is developed as a discretized form of the continuous Weibull distribution \citep{chakraborty15}. Namely, by defining the cumulative distribution function (cdf) of a continuous Weibull distribution by
\begin{equation*} 
	\label{F_CW}
	F_{CW}(y; q, \beta ) = 1 - \exp \Big[ - \Big( \frac{y}{(-\log q)^{-\frac{1}{\beta}}}\Big)^{\beta} \Big], \qquad y \ge 0,
\end{equation*} 
one can easily show that the probability mass function of the discrete Weibull distribution, associated to the cdf in Equation \ref{eq:dwmarginals}, is given by
\[
f(y; q, \beta) = q^{y^\beta}-q^{(y+1)^\beta}= F_{CW}(y+1) - F_{CW}(y) = \int_{y}^{y+1} f_{CW}(t) \mbox{dt}   \qquad y = 0,1,2, \ldots
\]
This creates a one-to-one connection between the latent continuous Weibull space, with the same parameters as the discrete Weibull distribution, and the Gaussian space, as depicted schematically in Figure \ref{fig:dwcwlink}.
\end{enumerate}

As it is clear also from the figure, each discrete observation is linked to an interval in the continuous space. This is the case for copula models on discrete data in general and will require special attention when it comes to inference, as we will discuss more in details in the next section.

\subsection{Bayesian inference for a DW graphical model}
Inference for copula graphical models involves estimation of the marginals and of the network component. A copula formulation enables us to learn the marginals  separately from the dependence structure of the $p$ random variables.

We first concentrate on the marginal components, that is the estimation of the regression coefficients $\bm{\theta}_j$ and $\bm{\gamma}_j$, $j=1,\ldots,p$.  Given $n$ observations on component $Y_j$, denoted with the vector $\bm{y}_j$, and on the d-dimensional vector of covariates, stored in the $n \times d$ matrix $\bm{x}$ with $\bm{x}_i$ the vector corresponding to the $i^{th}$ row, the likelihood for component $j$ is given by
\begin{equation*}
L_j(\bm{y}_j, \bm{x} \ | \ \boldsymbol{\theta}_j, \boldsymbol{\gamma}_j  ) 
= \displaystyle\prod_{i = 1}^{n} \left[  \left( \dfrac{e^{\bm{x}^t_i \boldsymbol{\theta}_j }}{1 + e^{\bm{x}^t_i \boldsymbol{\theta}_j }} \right)^{y_{ij} ^ {(e^{\bm{x}^t_i \boldsymbol{\gamma}_j })}} -   \left( \dfrac{e^{\bm{x}^t_i \boldsymbol{\theta}_j }}{1 + e^{\bm{x}^t_i \boldsymbol{\theta}_j }} \right) ^{ (y_{ij} + 1) ^ {(e^{\bm{x}^t_i \boldsymbol{\gamma}_j })}}   \right],
\end{equation*} 
where we consider the logit and log links on the $q$ and $\beta$ parameters, respectively. Based on this likelihood, we perform inference on the marginal components using an adaptive Metropolis-Hastings scheme, as in \cite{haselimashhadi18}. For the simulations and real data analysis in this paper, we set standard Gaussian priors on the regression coefficients $\bm{\theta}_j$ and $\bm{\gamma}_j$.

Once the marginals are estimated, inference of the network component requires an inverse mapping from the observed to the latent Gaussian space. As depicted visually in Figure \ref{fig:dwcwlink}, each observed discrete value corresponds to an interval in the latent Gaussian space with the same associated probability. Formally, 
given the $n \times p$ observed data $\bm{y}$ and the fitted marginals, the Gaussian latent variables $\bm{z}$ are constrained in the intervals
\begin{equation}
\mathcal{D}_F(\bm{y})  = \big \{ \bm{z} \in R^{n \times p} :  \Phi^{-1} \big(F_{ij}(y_{ij}-1) \big) < z_{ij} \le \Phi^{-1} \big(F_{ij}(y_{ij}) \big) \big \},
\label{eq:DWinterval}
\end{equation}
where we indicate with $F_{ij}$ the cdf of $Y_j$ when $\bm{X}=\bm{x}_i$.
Rather than condensing these intervals into a single point, as in \cite{cougoul19}, we retain this information within the MCMC sampling scheme, similar to the approach of \cite{dobra11} and \cite{mohammadi17} for ordinal data.

In particular, the extended rank likelihood function for a given graph $G$ and associated precision matrix $\bm{K}= {\bm R}^{-1}$ is defined as
\begin{equation*}
	\label{fomula:likelihood_graph}
	L_E(\bm{z} \in \mathcal{D}_F(\bm{y}); \bm{K} , G ) = \int_{\mathcal{D}_F(\bm{y})} P( \bm{z} \ | \ \bm{K}, G ) \ d\bm{z}
\end{equation*}
where $P(\bm{z} | \bm{K}, G)$ is the profile likelihood in the Gaussian latent space:
\begin{equation*}
 P(\bm{z} | \bm{K}, G) \propto |\bm{K}|^{n/2} \exp \biggl\{-\frac{1}{2} \mbox{Tr}(\bm{K}\mathbf{U}) \biggr\}
 \end{equation*}
with $\bm{U} = \bm{z}^{t} \bm{z}$ the sample moment. 
The likelihood is combined to priors to lead to the posterior
\begin{equation}
\label{eq:posterior_graph}	
 P\big( \bm{K}, G  \ | \ \bm{z} \in \mathcal{D}_F(\bm{y}) \big) \propto L_E( \bm{z} \in \mathcal{D}_F(\bm{y});\bm{K} , G ) \ P(\bm{K} \ | \ G) \ P(G)
\end{equation}
where  $P(\bm{K} \ | \ G)$ denotes the prior distribution on the precision matrix $\bm{K}$ for a given graph structure $G$ and $P(G)$ denotes a prior distribution for the graph $G$. Similar to \cite{mohammadi17}, one can show that the posterior distribution of each marginal $Z_j$ conditional on the other $\bm{Z}$ and on the precision matrix $\bm{K}$ is given by a Gaussian distribution
  $$Z_j | \bm{K},\bm{Z}_{V \setminus \{ j \} } = \bm{z} \sim N \Big(- \sum_{k} { \dfrac{K_{jk} z_{k} }{K_{jj}} }, \dfrac{1}{K_{jj}}\Big),$$
truncated on the interval
\[
\Big(\Phi^{-1} \big(F_{j}(Y_{j}-1) \big) < Z_{j} \le \Phi^{-1} \big(F_{j}(Y_{j})\big)\Big],
\]
with $F_j(\cdot)$  the discrete Weibull cdf linking $Y_{j}$ to $\bm{x}$.

As regards to the prior specification on the graph $G$, we consider an Erd\"{o}s-R\'{e}nyi random graph with a  prior probability of a link $\pi \in (0,1)$ (which we set to 0.2, representing the case of a sparse graph, unless stated otherwise). For other options, see \cite{dobra11} and \cite{mohammadi15}. As for the precision matrix $\bm{K}$, conditional on a given graph $G$,  we consider a G-Wishart distribution, defined by 
\[
P(\bm{K}|G)=\frac{1}{I_G (b,\bm{D})} |\bm{K}|^{(b-2)/2} \exp \left\{ -\frac{1}{2} \mbox{Tr}(\bm{D}\bm{K}) \right\},
\]
where $b > 2$ is the degree of freedom, $\bm{D}$ is a symmetric positive definite matrix, and $I_G (b,\bm{D})$ is a normalizing constant \citep{roverato02}.  For the simulations and real data analysis in this paper, we set  $b=3$ and $\bm{D}=\mathbb{I}_p$, following \cite{mohammadi21}. 

As the space of possible graphs is very large, computationally efficient search algorithms are needed to sample from the posterior distribution (\ref{eq:posterior_graph}). To efficiently explore the graph space, we consider the birth-death Markov chain Monte Carlo (BDMCMC) search algorithm developed by \cite{mohammadi15}. In particular, the algorithm explores the graph space by either adding (birth) or deleting (death) an edge to a graph $G=(V,E)$, independently of the rest and via a Poisson process with birth/death rates given by
\begin{equation}
\label{eq:rate}
R_e( G, \bm{K} ) = \min \left\{ \frac{P(G^{*},\bm{K}^{*} |\bm{z})}{P(G,\bm{K} |\bm{z})}, 1 \right\},  \text{ for each  } e \in \{ E \cup \eb \},
\end{equation}
where $G^*=(V, E \cup \{e\} )$ for the birth of an edge $e \in \eb$, while $G^* = ( V, E \setminus \{e\} )$ for the death of an edge $e \in E$, and $\bm{K}^{*}$ is the corresponding precision matrix.
Since the birth/death events are independent Poisson processes, the time between two successive events has a mean waiting time given by
\begin{equation}
\label{eq:waiting time}
W( G, \bm{K}) = \frac{1}{\sum_{e  \in \{ E \cup \eb \} }R_e( G, \bm{K} ) }.
\end{equation} 
Based on the above birth/death rates and waiting times, the birth and death probabilities that govern the move to a new graph are given by
\begin{equation}
\label{eq:prop rate}
P(\mbox{birth/death of edge  } e \in \{ E \cup \eb \} ) = R_e( G, \bm{K} )\times W( G, \bm{K}).
\end{equation}

\begin{figure}[!h]
\begin{center}
 \includegraphics[width=5.3in]{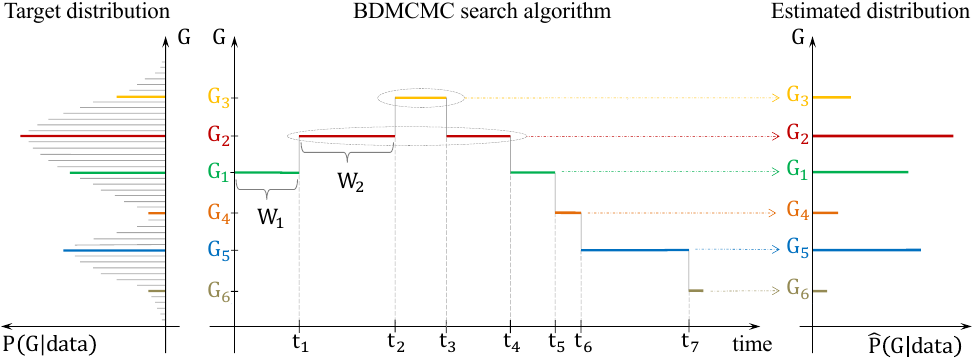} 
\end{center}
    \caption{ 
 Graphical representation of the BDMCMC search algorithm over the graph space for the Step 2 of Algorithm \ref{alg:BDMCMC}.
The left panel shows the target posterior distribution of the graphs, while the right panel represents its estimation based on the total waiting times of the graphs visited by the algorithm.      
The middle panel visualizes how the algorithm explores the graph space, where $\left\{ W_1, W_2,... \right\}$ are the waiting times and $\left\{ t_1, t_2,... \right\}$ are the jumping times of the algorithm. This figure is adapted from \cite{mohammadi21}. \label{fig:bdmcmc} }
\end{figure}

The pseudo-code for the BDMCMC search algorithm for sampling from the target posterior distribution (\ref{eq:posterior_graph}) is reported in  Algorithm \ref{alg:BDMCMC}.
\begin{algorithm} 
\caption{\label{alg:BDMCMC} BDMCMC search algorithm for \texttt{GCGM} with DW marginals}
 \KwIn{ A graph $G = (V,E)$ with a precision matrix $\bm{K}$ and data $\bm{y}$ and $\bm{x}$. }
 \For{$N$ iteration}{
 \textbf{Step 1}: Sample the latent data for each marginal $j$, updating the latent $n$ values $\bm{z}_j$ from their full conditional distribution:
  $$Z_j | \bm{K},\bm{Z}_{V \setminus \{ j \} } = \bm{z} \sim N \Big(- \sum_{k} { \dfrac{K_{jk} z_{k} }{K_{jj}} }, \dfrac{1}{K_{jj}}\Big),$$
each truncated on its corresponding interval in $\mathcal{D}_{F_j}(\bm{y})$ from  Equation (\ref{eq:DWinterval})\;
  \textbf{Step 2}:  \For{all the possible jumps in parallel}{
   Compute the birth and death rates by Equation \ref{eq:rate}\;
   }
   Compute the waiting time by Equation \ref{eq:waiting time}\;
   Sample the graph based on the birth/death probabilities in Equation \ref{eq:prop rate}\;
   \textbf{Step 3}: Sample the precision matrix, according to the updated graph\;
 }
 \KwOut{ Samples from the target posterior distribution \eqref{eq:posterior_graph}. } 
\end{algorithm}
The first step of Algorithm \ref{alg:BDMCMC} is to update the latent variables given the observed data. 
Then, in step 2, on the basis of the sampled latent data, the algorithm computes the birth/death rates. This is done in parallel since the rates associated to each edge can be calculated independently of each other. For details on how to calculate the birth/death rates see \cite[Section 2]{mohammadi21}, while Figure \ref{fig:bdmcmc} provides a visualization of the algorithm.
Finally, step 3 of the algorithm can be done by exact sampling from a G-Wishart distribution, as in \cite{lenkoski13}. 

Following from the Bayesian inference of the Gaussian copula graphical model with discrete Weibull marginals, one can extract any information of interest for the analysis. In particular, from the marginal components, one obtains the posterior distribution of the regression coefficients and can investigate any effect of interest, while from the graph posterior, one can calculate the posterior edge inclusion probabilities:
\begin{equation}
P(\text{edge e} \in E~|\text{data}) = \frac{\sum_{t=1}^{N} 1(e \in G^{(t)}) W(G^{(t)},\bm{K}^{(t)})}{\sum_{t=1}^{N} W(G^{(t)},\bm{K}^{(t)})},
\label{eq:edgeprob}
\end{equation}
where $N$ denotes the MCMC iterations (after burn-in) and $1(e \in G^{(t)})=1$ if $e \in G^{(t)}$ and zero otherwise. These probabilities capture the full uncertainty on the graph learning, which is particularly useful in high dimensional settings such as the micriobiome data.

\section{Simulation study}
\label{sec:simulation}

The main objective of the proposed method is that of learning the underlying structure of dependence from complex and heterogeneous count data, which are routinely generated in genomic studies. We therefore conduct a simulation study to measure the performance of the method in this setting. 

For the simulations, we consider networks with $p$ nodes and a sparse random graph structure. 
Given a graph $G$ and marginals $F_j(\cdot)$, $j=1,\ldots,p$, we use the following procedure to simulate count data. We first generate a precision matrix from a G-Wishart distribution with $b=3$ and $\bm{D}=\mathbb{I}_p$, and standardize it to the inverse of a correlation matrix. We then draw $n$ multivariate normal samples from $N_p(\bm{0}, \bm{K}^{-1})$. This generates a matrix $\bm{z}$ of dimension $n \times p$. Finally, we obtain the discrete data using $y_{ij} = F_j^{-1}(\Phi(z_{ij}))$, for $i=1,\ldots,n$ and $j=1,\ldots,p$, with $\Phi(\cdot)$ the standard normal distribution and $F_j(\cdot)$ a distribution function of a specified shape as detailed in each study below.

We evaluate the performance of the method in terms of parameter estimation and graph recovery. For the first, we compare the true precision matrix $\bm{K}$ with the posterior mean estimate $\widehat{\bm{K}}$ using the Kullback-Leibler divergence
\[
KL(\widehat{\bm{K}})= 0.5(\mbox{Tr}(\bm{K}\widehat{\bm{K}}^{-1}) + \mbox{Tr}(\widehat{\bm{K}}\bm{K}^{-1}))-p,
\]
while, for graph recovery, we use the function \texttt{auc} in the \texttt{pROC} R package to calculate the area under the Receiver Operating Characteristic (ROC) curve. The latter is obtained by setting cutoffs on the posterior edge inclusion probabilities in Equation (\ref{eq:edgeprob}). At each cutoff,  the $x$ and $y$ coordinates of the point on the curve are given by the false positive rate and the true positive rate, respectively, of the estimated graph for that cutoff and with respect to the true graph structure.

\subsection{Effect of increasing p and graph density}
In a first simulation study, we evaluate the performance of the proposed  Gaussian copula graphical model with discrete Weibull marginals, which we abbreviate to \texttt{DWGM}. In particular, we test how the performance of the method is affected by the dimensionality $p$ and the sparsity of the graph $G$.

We generate data as described before, with marginals  $F_j(\cdot)$ given by discrete Weibull distributions linked to one external covariate. In particular, we consider a binary covariate $X$ drawn from a Bernoulli(0.5), e.g., observations split into two groups (like the two environments, stool and saliva, in the real application in Section \ref{sec:realdata}). For the regression parameters in Equation (\ref{dw_linkfun}), we set a constant $\log(\beta(x))= \gamma_0 = \log(0.7)$ for all $p$ variables, while $q$ values that differ across the two conditions and the $p$ variables are obtained by setting $\log(q(x)/(1-q(x))= \theta_0+\theta_1x$, with  $\theta_0$ drawn from a N(0,0.1) and $\theta_1$ from a N(2,0.01). This choice of parameters leads to generally over-dispersed data \citep{peluso19}.

\begin{figure}[!h]
\begin{center}
\includegraphics[scale=0.42]{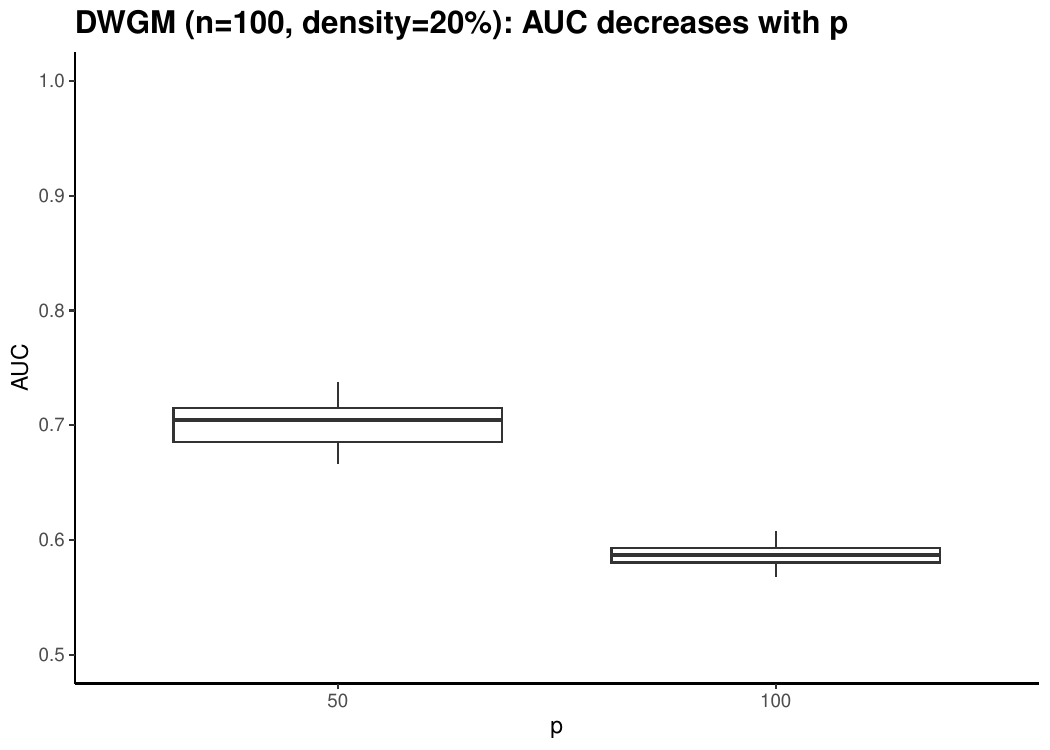}
\includegraphics[scale=0.42]{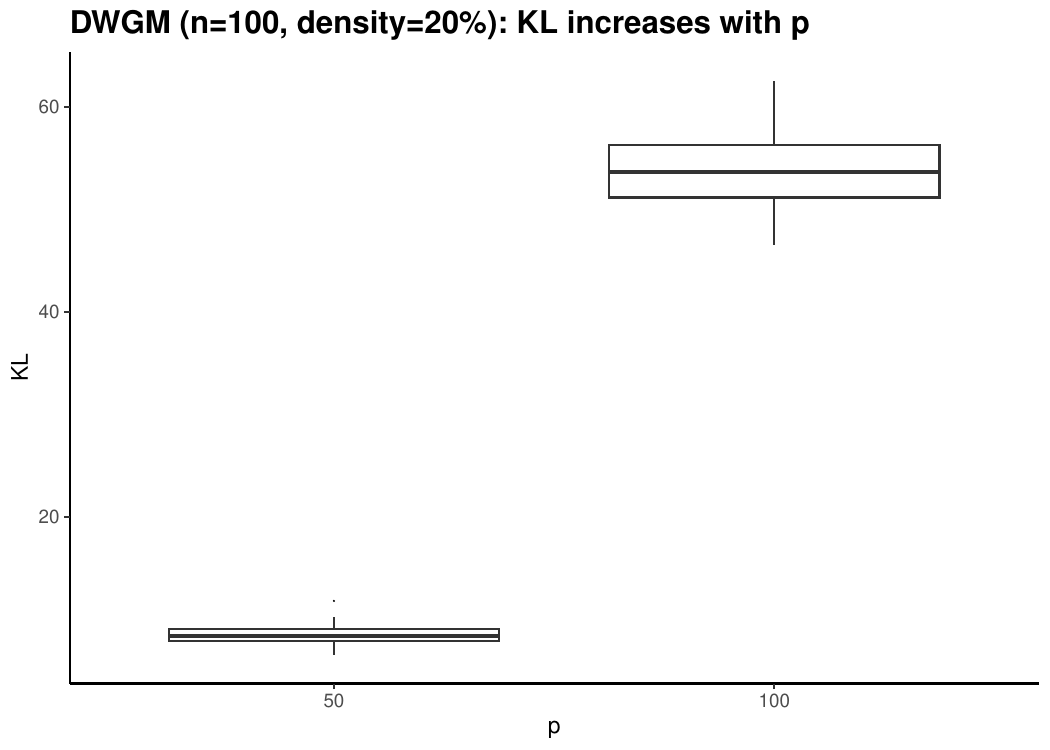}
\includegraphics[scale=0.42]{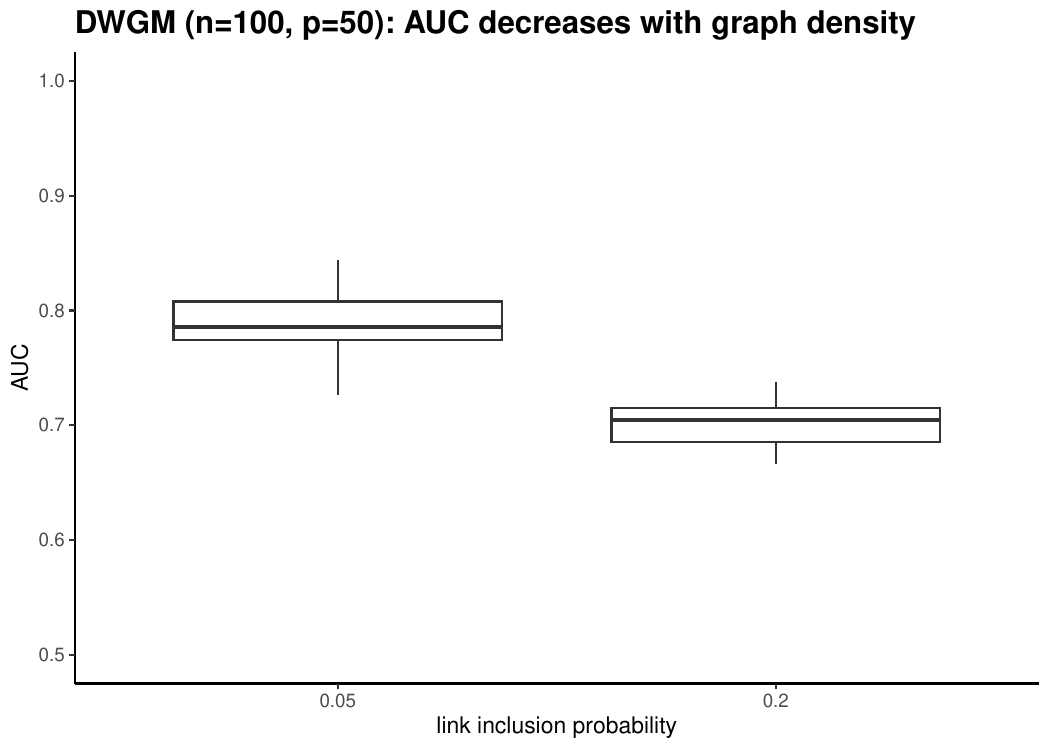}
\includegraphics[scale=0.42]{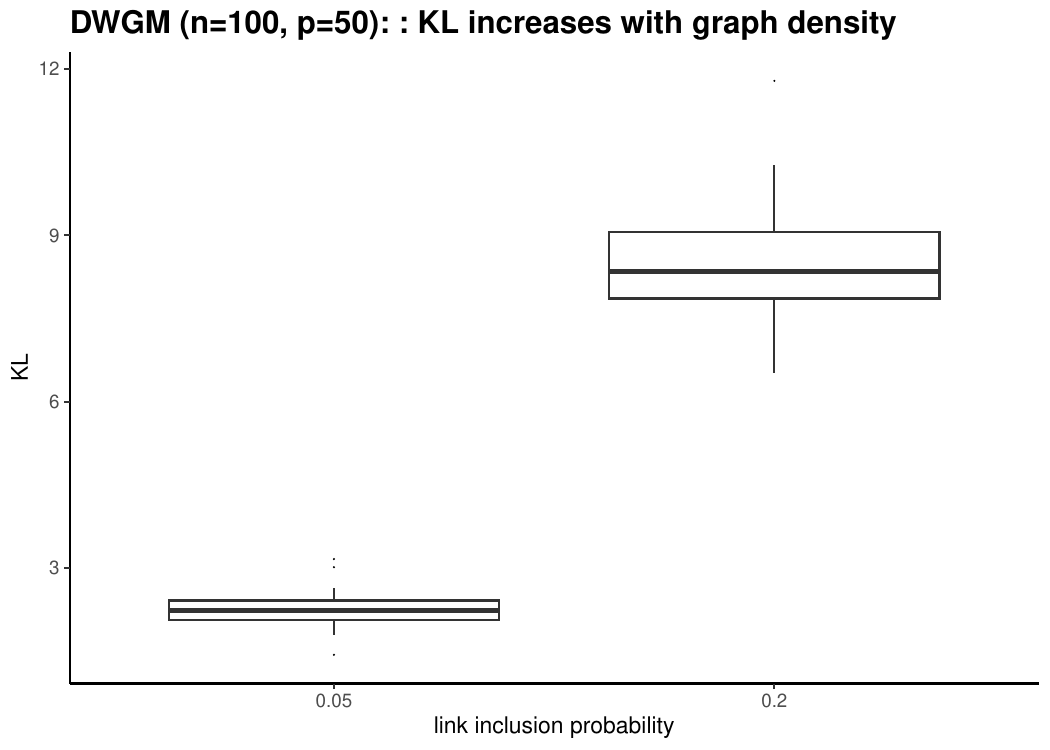}
\end{center}
\caption{Performance of DWGM, in terms of accuracy of network recovery (left) and parameter estimation (right), decreases with increasing dimensions (top) and graph density (bottom). Data are simulated from a Gaussian copula graphical model with DW marginals. Boxplots are the result of 50 simulations. \label{fig:sim1}}
\end{figure}

We set $n=100$, $p=50$ and a link inclusion probability in $G$ equal to $0.05$. We then measure how the performance of the method varies when increasing $p$ to $100$ or the graph density to $20\%$. Figure \ref{fig:sim1} shows the AUC and KL values across 50 simulated datasets. For each dataset, we run the Bayesian inferential procedures for 1000 iterations for each marginal and 10k iterations for the structure learning (Algorithm \ref{alg:BDMCMC}), and set the priors as specified in the description of the method. 
As expected, and in line with similar studies in the literature, we find that the performance deteriorates as $p$ increases, both in terms of graph recovery (Figure \ref{fig:sim1}, top left) and parameter estimation (Figure \ref{fig:sim1}, top right). Similarly, we find a deterioration in both performance measures as the link inclusion probability increases from $0.05$ to $0.2$, i.e., the denser the graph becomes (Figure \ref{fig:sim1}, bottom panel).

The computational time is mostly affected by $p$. In particular, an analysis on one of the datasets with $p=50$, $n=100$ and link inclusion probability equal to $0.2$ required approximately $80$ seconds, compared to $333$ seconds when $p=100$.

\subsection{Effect of covariate adjustment on structural learning}
In a second simulation study, we evaluate the effect of covariate adjustment on structural learning. To this end, we consider the same generative process as in the previous section and concentrate on the case $p=50$, $n=100$ and $0.2$ link inclusion probability. As a benchmark, we also consider the case of no covariates in the marginal models, i.e., $\theta_1=0$. We then compare the proposed \texttt{DWGM} with the Gaussian copula graphical model (\texttt{GCGM}) for ordinal data of \cite{mohammadi17}, implemented in the R package {\tt BDgraph}. We use 10k iterations also in this case and the same prior specifications.
 In the absence of covariates, the \texttt{GCGM} method is similar to our approach, the only difference being the (non-parametric) empirical marginal distributions used in \texttt{GCGM} versus the parametric distributions with an unbounded support for the marginals in \texttt{DWGM}.  Clearly, being non-parametric, \texttt{GCGM} does not lend itself easily to the inclusion of external covariates in the marginals. 

\begin{figure}[!h]
\begin{center}
\includegraphics[scale=0.42]{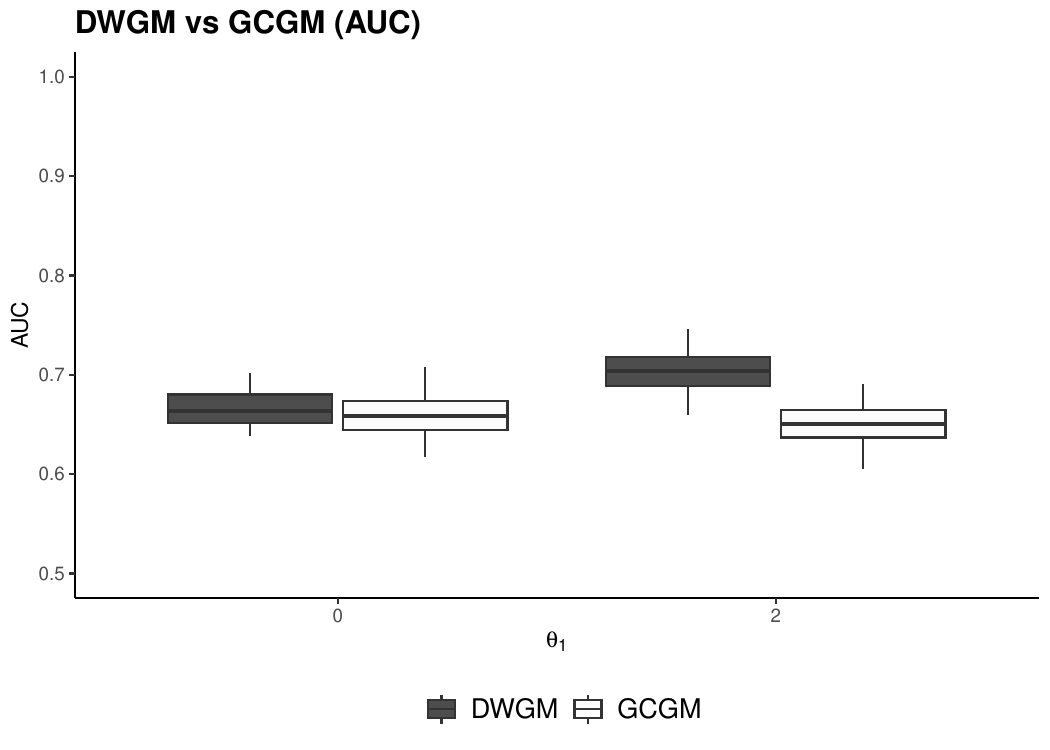}
\includegraphics[scale=0.42]{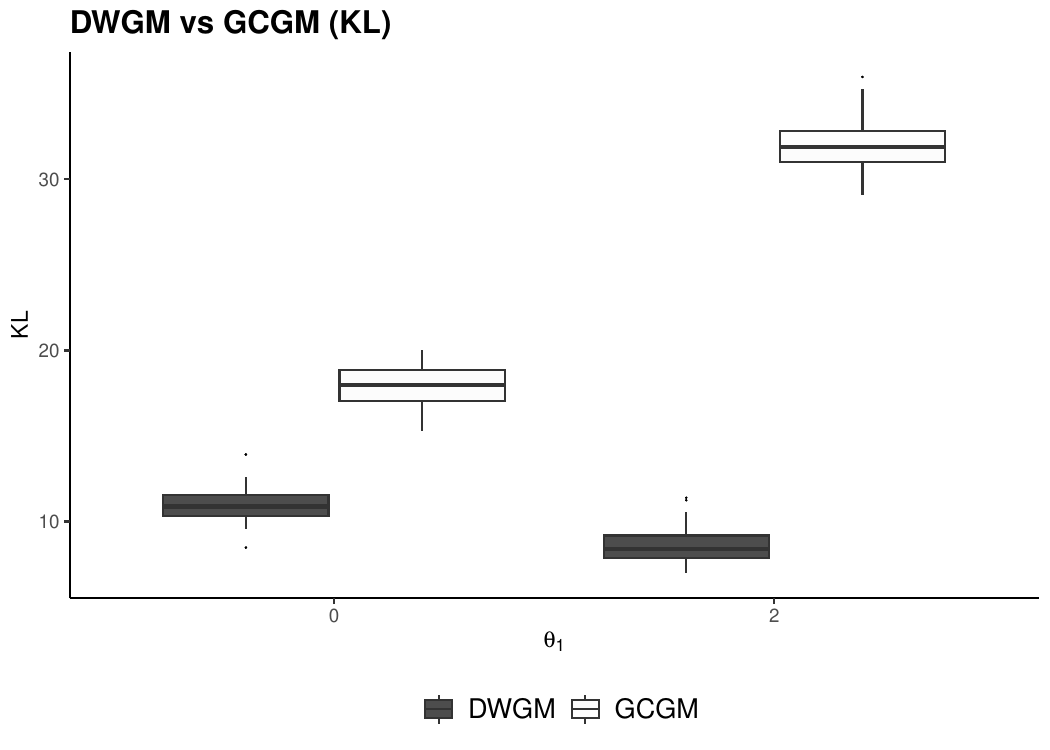}
\end{center}
\caption{Performance of DWGM, in terms of accuracy of network recovery (left) and parameter estimation (right), is compared with a Gaussian copula graphical model with non-parametric marginals (\texttt{GCGM}). Data are simulated from a Gaussian copula graphical model with $n=100$, $p=50$, a link inclusion probability of $0.2$ and DW marginals without ($\theta_1=0$) and with ($\theta_1=2$, on average across the $p$ variables) the presence of an external covariate $X$. Boxplots are the result of 50 simulations.\label{fig:sim2}}
\end{figure}

Figure \ref{fig:sim2} evaluates the two approaches in terms of graph recovery (left) and parameter estimation (right) across $50$ simulations.
As expected, in the absence of covariates ($\theta_1 = 0$), the two approaches perform similarly, although the correctly specified parametric marginals lead to a better estimation of the precision matrix.  In contrast to this, the improvement in structural learning after adjusting for covariate effects is clear when data are simulated with a $\theta_1$ marginal effect equal to $2$ on average across the $p$ variables, both in terms of graph recovery and parameter estimation. Looking closely at graph recovery, and estimating a graph structure by setting a $0.5$ cutoff on the posterior edge probabilities, we find that the \texttt{GCGM} estimated graph is in general denser (14\% density on average versus 9\% for \texttt{DWGM}), with a worse false positive rate (9\% on average versus 3\% for \texttt{DWGM}) and only a marginally better false negative rate (66\% versus 67\% for  \texttt{DWGM}).
This suggests that not correcting for the covariates at the marginal level may result in the detection of spurious dependences between the variables, as noted also in other studies \citep{vinciotti16, vinciotti23}. 

The computational time for one of the datasets was about 79 seconds for \texttt{DWGM} and 36 seconds for  \texttt{GCGM}. The difference is mostly attributed to the fitting of the marginal distributions. This has a negligible computational time in \texttt{GCGM}, while  MCMC sampling for the marginal fitting in \texttt{DWGM} required about 45 seconds. Future versions of the \texttt{BDgraph} package will consider a parallel implementation for the marginal fitting across the $p$ variables.

\subsection{Effect of miss-specified marginal distributions}
In a third simulation study, we evaluate the robustness of \texttt{DWGM} to count data simulated with negative Binomial marginals. We consider a similar setting to the first simulation, i.e., one binary covariate,  a constant  dispersion parameter $\phi=0.5$ and a mean $\mu$ dependent on $X$, with $\log(\mu(x) = \theta_0+\theta_1x$, with  $\theta_0$ drawn from a N(0,0.1) and $\theta_1$ from a N(2,0.01) across the $p$ variables. As with the second simulation, we fix $n=100$, $p=50$ and a link inclusion probability equal to $0.2$.

We compare our proposed method with a Gaussian copula graphical model with negative Binomial regression marginals, implemented in the R package \texttt{rMAGMA} \citep{cougoul19}.  As well as using a different distribution for the marginals, inference in \texttt{rMAGMA} is conducted using a frequentist paradigm and, in addition, it does not make use of the extended rank likelihood approach.  Indeed, the fitted marginals are used to transform the data into the latent variables by taking the mean of the interval, i.e., $z_{ij}=\Phi^{-1}\Big(\dfrac{F_j(y_{ij}-1)+F_j(y_{ij})}{2}\Big)$, and then graphical lasso is used on the transformed data.

\begin{figure}[!h]
\begin{center}
\includegraphics[scale=0.42]{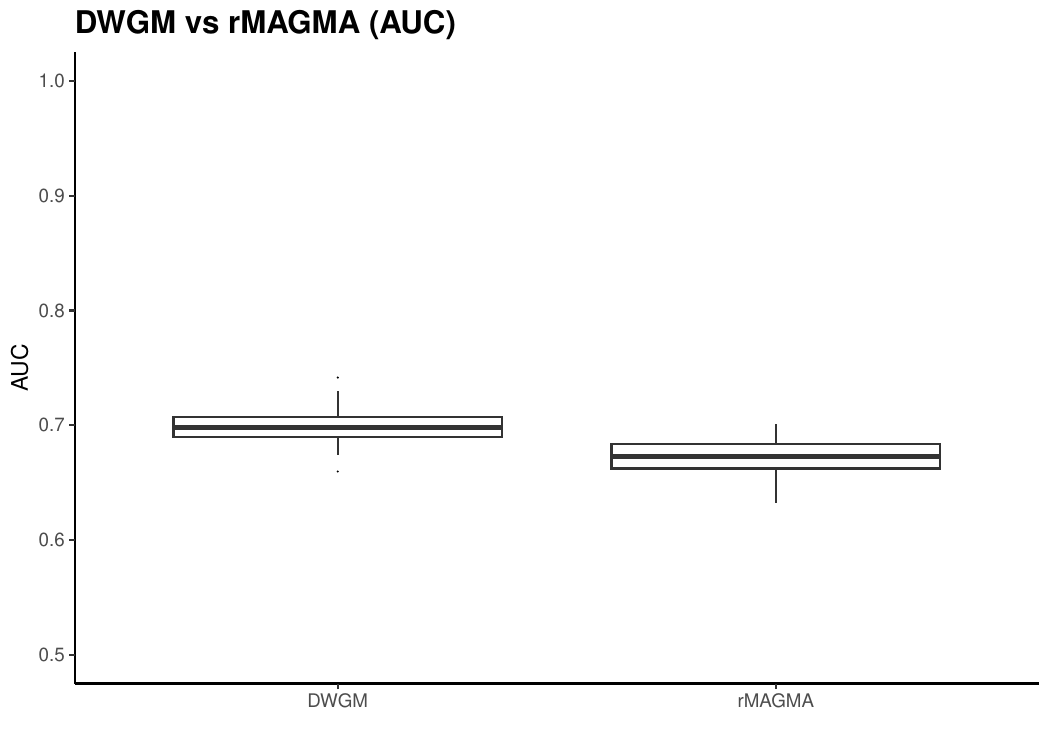}
\includegraphics[scale=0.42]{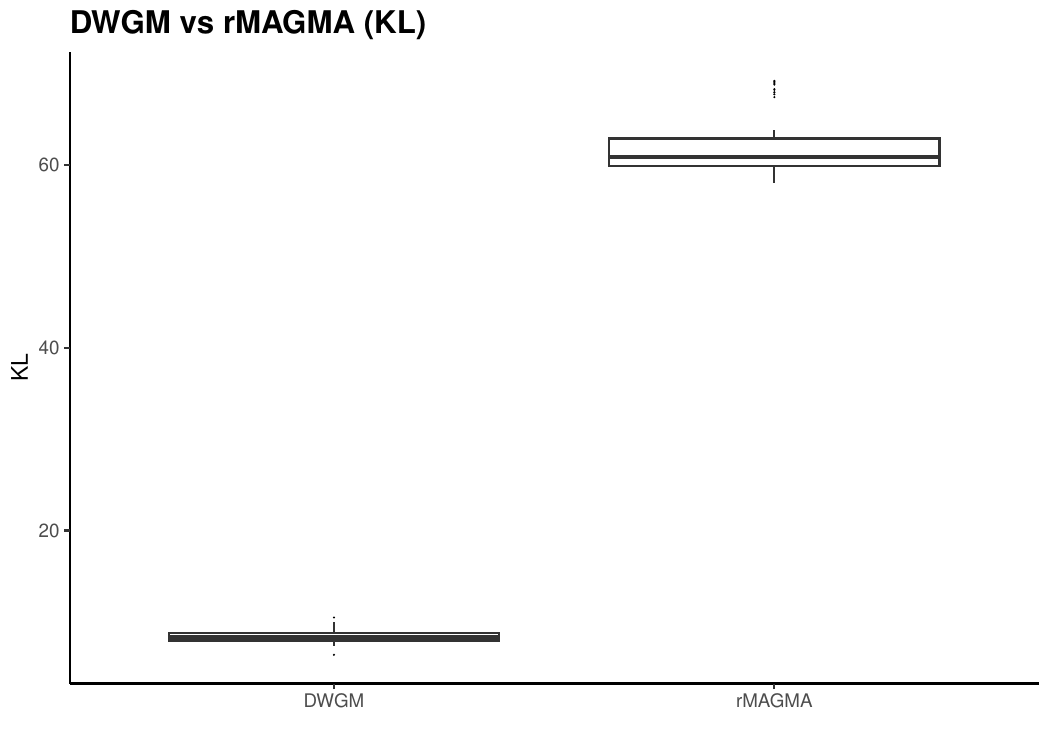}
\end{center}
\caption{Performance of DWGM, in terms of accuracy of network recovery (left) and parameter estimation (right), is compared with a Gaussian copula graphical model with parametric NB marginals (\texttt{rMAGMA}). Data are simulated from a Gaussian copula graphical model with $n=100$, $p=50$, a link inclusion probability of $0.2$ and NB marginals. Boxplots are the result of 50 simulations. \label{fig:sim3}}
\end{figure}

Figure \ref{fig:sim3} reports the accuracy of the methods in terms of graph recovery (left) and parameter estimation (right). For \texttt{rMAGMA}, which is based on penalised inference,  the ROC curve is constructed across the path of solutions generated by the tuning penalty parameter, while the Kullback-Leibler is calculated based on the optimal model selected using the stability approach for regulation selection (stars) criterion \citep{liu10}.

The results show, firstly, similar results for \texttt{DWGM} compared to the previous simulations (Figure \ref{fig:sim1} and Figure \ref{fig:sim2} for $p=50$, $n=100$ and 20\% graph density), suggesting a robustness of \texttt{DWGM} against the marginal model specification. Secondly, \texttt{DWGM} has a better performance than  \texttt{rMAGMA}, particularly when it comes to the accuracy of the estimated precision matrix $\bm{K}$ of the optimal model. Since \texttt{rMAGMA} is in fact correctly specified in this simulation,  we speculate that the posterior edge probabilities of a Bayesian structural learning procedure lead to a better separation between the presence/absence of links than the edge weights calculated across the penalised path of the frequentist \texttt{rMAGMA} approach.

\section{Inferring the network of the microbiota}
\label{sec:realdata}
In this section, we use Gaussian copula graphical models with discrete Weibull marginals to recover the network of interactions between microbial species.  Interactions between microbes are fundamental in shaping the structure and functioning of the human microbiota, and their malfunctioning has been linked to a number of medical conditions. A lack of understanding of how these interactions shape and evolve
makes it difficult to predict their relevance in biomedical fields. For these reasons, microbiota systems have been intensively studied in recent years. Large consortia have developed technologies for the collection of high-throughput data of the microbiome, e.g., the Human Microbiome Project \citep{HMP} and the Metagenomics of the Human Intestinal Tract (MetaHIT) project \citep{qin10}. These have paved the way for further studies investigating the association of the micriobiota functioning with a number of medical conditions, such as obesity \citep{lechatelier13} and diabetes \citep{pedersen16}, as well as with the response to certain treatments, such as immunotherapy \citep{lee22}.

In this illustration, we focus in particular on the gut and oral microbiome.
The microbial communities in the mouth and colon are connected anatomically via the saliva. However, the extent to which oral microbes reach and colonize the gut is yet under debate \citep{rashidi21}. To resolve this long-standing controversy, many studies have been devoted to study jointly the human stool and saliva microbiome profiles. To this end, we apply our methodology to recover a core network of interactions between microbes across the two different environments. Crucially, the method that we have developed takes into account both the fact that the OTU abundances may differ marginally between the body sites (stool and saliva) and that the data may be affected by potential experimental effects.

As in \cite{cougoul19}, we retrieve the 16S variable region V3-5 data from the Human Microbiome Project \citep{HMP} and perform the analysis at the level of Operating Taxonomic Units (OTUs). After filtering samples with less than 500 reads, we consider micriobiomes from 663 healthy individuals, with microbial concentration measured from either stool or saliva. We then restrict our attention to the 155 OTUs which are present in at least 25\% of the samples and with more than two distinct observed values in both the saliva and stool samples. 
Finally, in order to account for varying sequencing depths change significantly between samples,  we estimate the library size of each sample by the geometric mean of pairwise ratios of OTU abundances of that sample with all other samples \citep{cougoul19}. 

In the next sections, we use the proposed approach on the micriobiome data with $p=155$ OTUs (the nodes of the network) and $n=663$ samples, accounting for the marginal effects of the location in the body (stool or saliva) and the library size of each biological sample.

\subsection{Accounting for covariates via DW regression marginals}
 We fit discrete Weibull marginal models, linking both parameters $q$ and $\beta$ to body site and sequencing length. We take sequencing length in the log scale, which is more in line with its use in the literature as an offset of a negative Binomial model \citep{cougoul19}. Including also an interaction between the two covariates leads to 8 regression parameters  per marginal component.  As the data are sparse (with a percentage of zeros per OTU ranging from 40.7\% to 75\%), we fit also a zero-inflated discrete Weibull distribution, with a  zero inflation parameter $\pi_j(\bm x)$ for component $j$, which we let vary between the two body sites. On each of the two additional parameters, we place a Beta(1,1) prior distribution. 
 
 For each marginal and for each of the two models (discrete Weibull and zero-inflated discrete Weibull), we use 10k MCMC iterations, retaining the last 25\% as samples from the posterior distribution. Trace plots of the regression parameters showed that this number of iterations was sufficient to reach convergence. A comparison of the zero-inflated versus the standard DW regression model using the Bayesian Information Criterion (BIC) showed that 24 out of the 155 OTUs necessitated the zero-inflated component of the model.  
As a matter of comparison, we also fitted negative Binomial, using its most common formulation with a mean dependent on the covariates and a constant dispersion parameter, as in \cite{cougoul19}. Here we found that 15 out of the 155 OTUs were better fitted with a zero-inflated NB model. These results show how using a zero-inflated model upfront because the data are very sparse, as done in most of the literature on micriobiome analyses, may not necessarily be the best option and that conducting model selection between the two models, as done in this paper, is a better choice. 

\begin{figure}[!h]
\begin{center}
\includegraphics[scale=0.4]{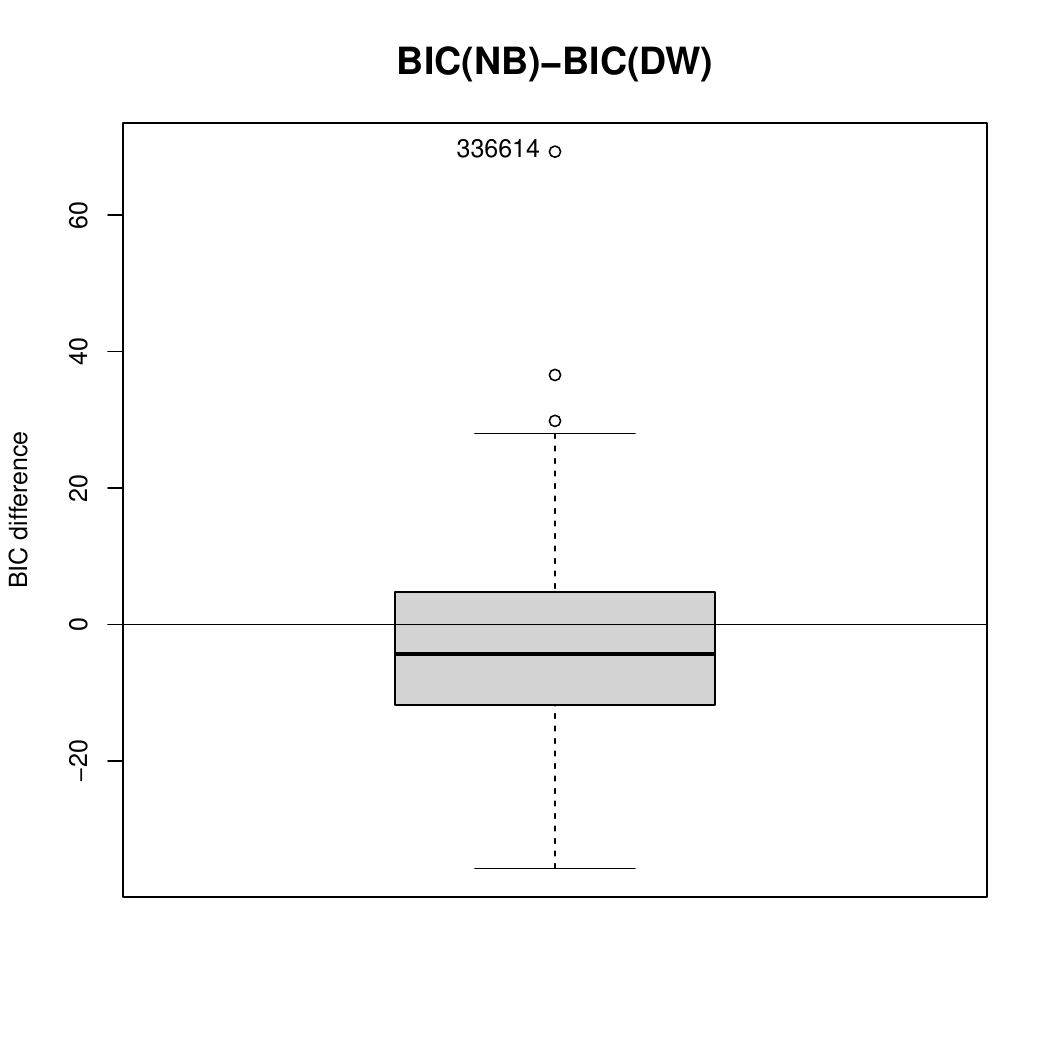}
\includegraphics[scale=0.4]{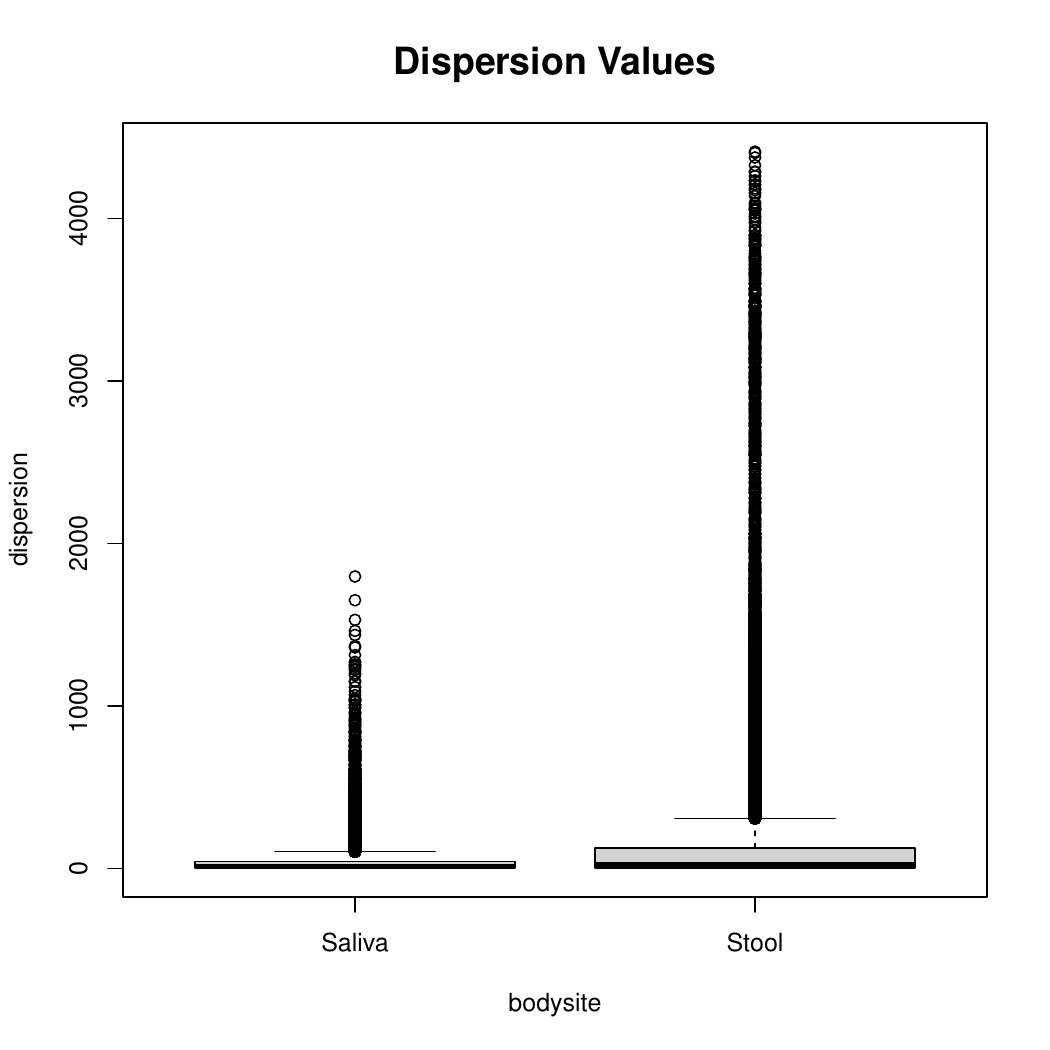}\\
\includegraphics[scale=0.4]{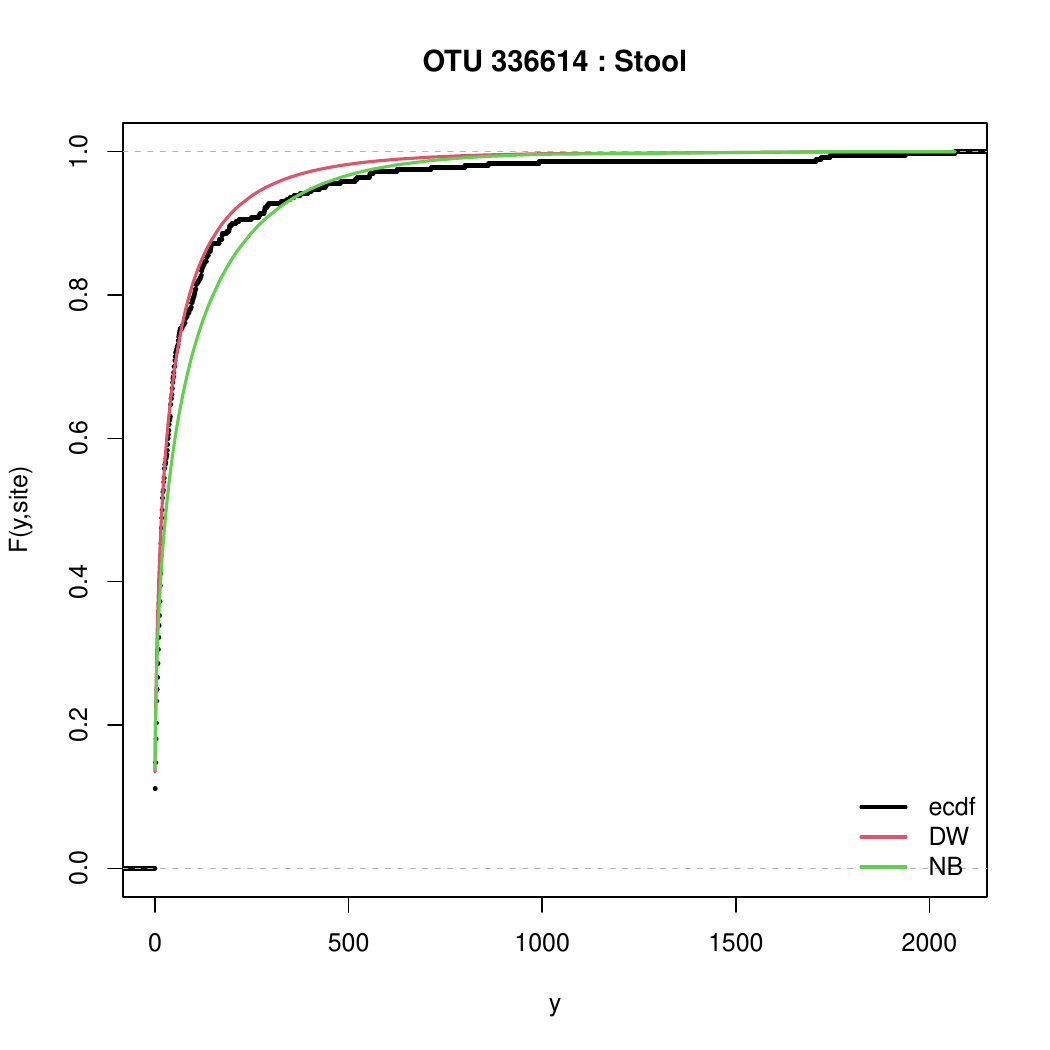}
\includegraphics[scale=0.4]{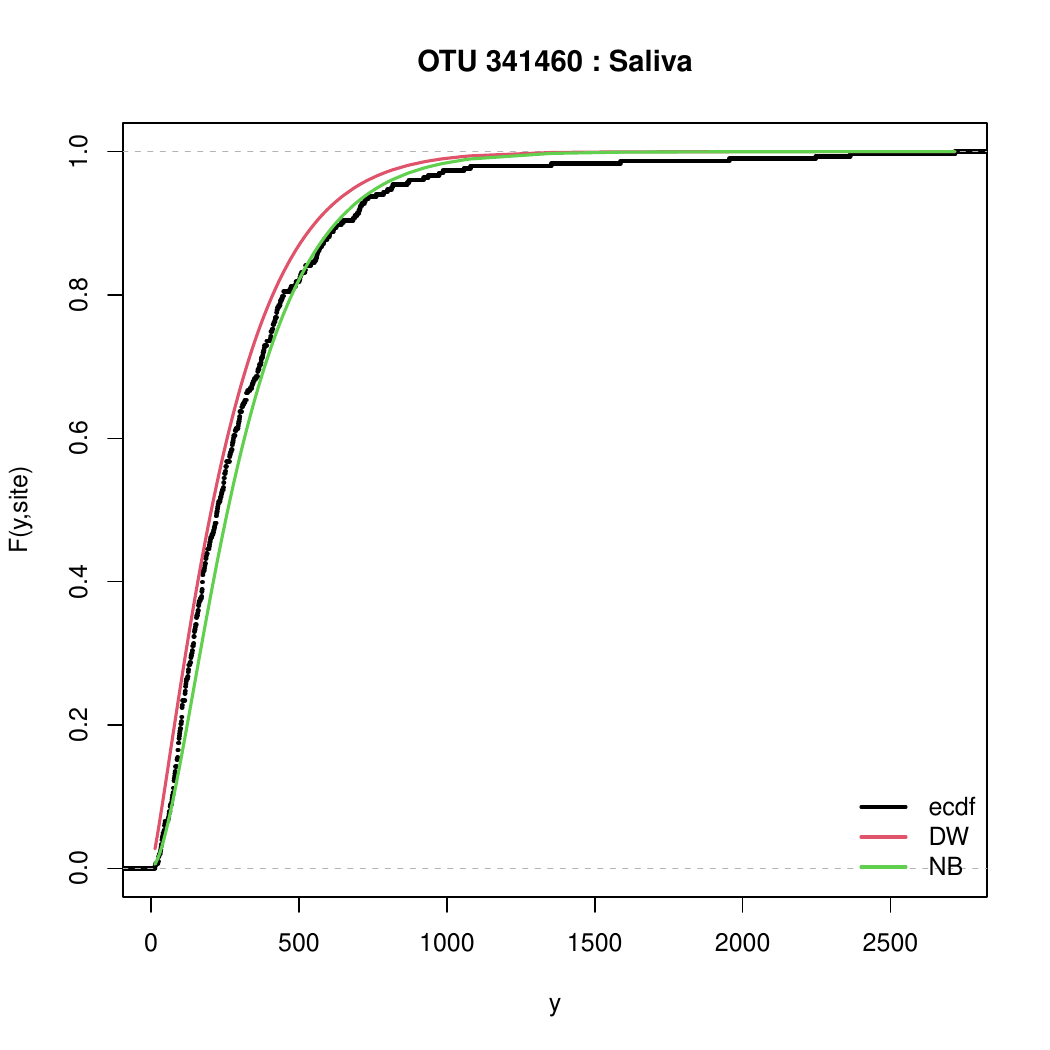}
\end{center}
\caption{Top: Boxplot of BIC differences between the NB model and the DW model across all 155 OTUs (left) and dispersion levels from the fitted DW model for each OTU and each observation, split by body site (right). Bottom: Cumulative distribution functions (empirical and fitted) corresponding to a specific body site for the OTU best fitted by DW (BIC difference = 68.26235, left) and by NB (BIC difference = -31.07967, right) .\label{fig:bicdiff}}
\end{figure}

Figure \ref{fig:bicdiff} reports the results, whereby for each OTU we consider  the best model between the zero-inflated and the non-zero inflated version. The BIC comparison shows similar performance between discrete Weibull and negative Binomial models, with some cases showing a significantly better fit for discrete Weibull. 
The top right plot shows the dispersion ratio (i.e., the variance divided by the mean) from the fitted DW marginals for each observation and each OTU. The plot shows how the data are highly over-dispersed in both conditions, a setting where NB is typically the default choice. Finally, the bottom plots show the OTU with the largest BIC difference (left plot), i.e., the OTU best fitted by DW when compared to NB,  and the OTU with the lowest BIC difference (right plot). In both cases, we plot the cumulative distribution functions of DW and NB associated to the body site which shows the biggest difference, while taking an average of the parameters across the normalizing factor. Superimposing these fitted distributions on the empirical cumulative distribution functions associated to the two groups shows the extent of the discrepancy between the two models. 

\begin{figure}[!h]
\begin{center}
\includegraphics[scale=0.4]{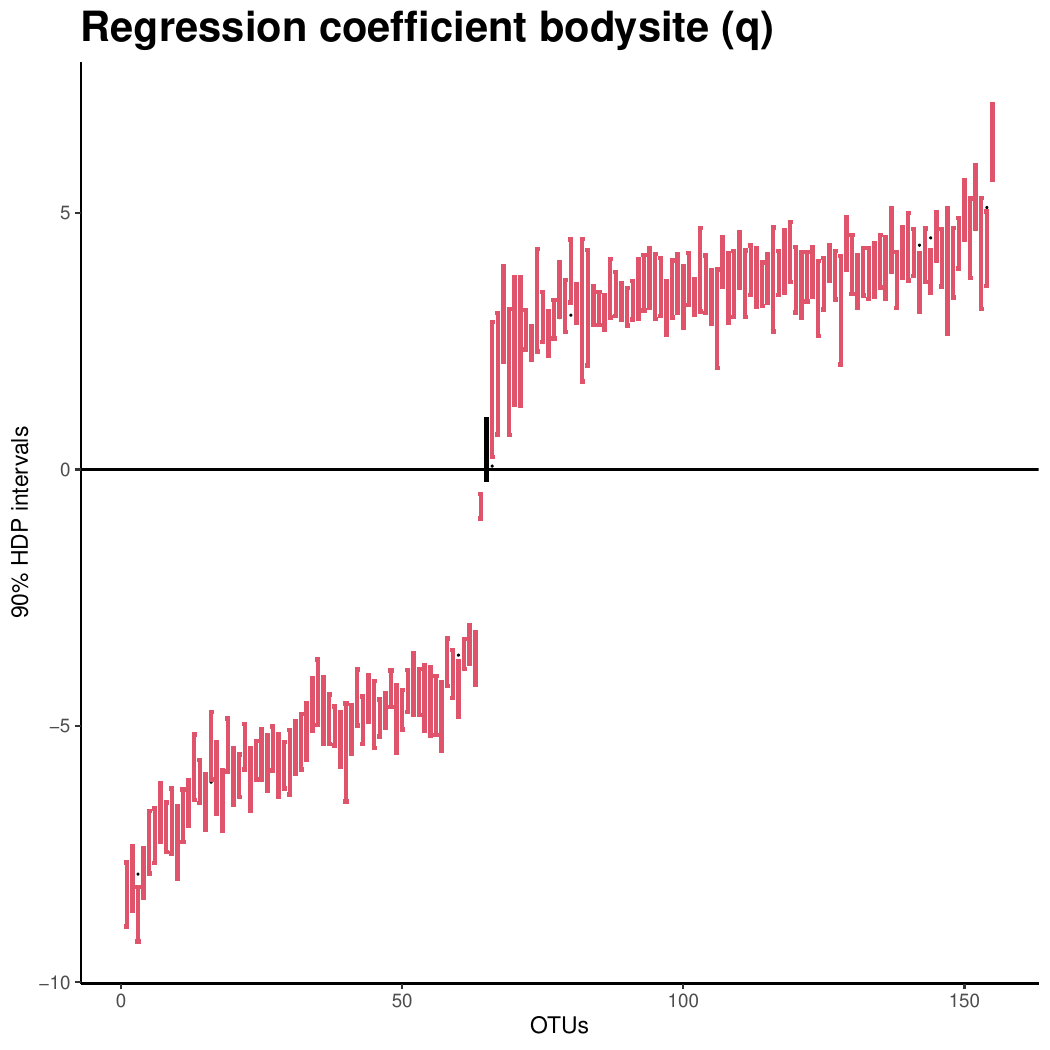} 
\includegraphics[scale=0.4]{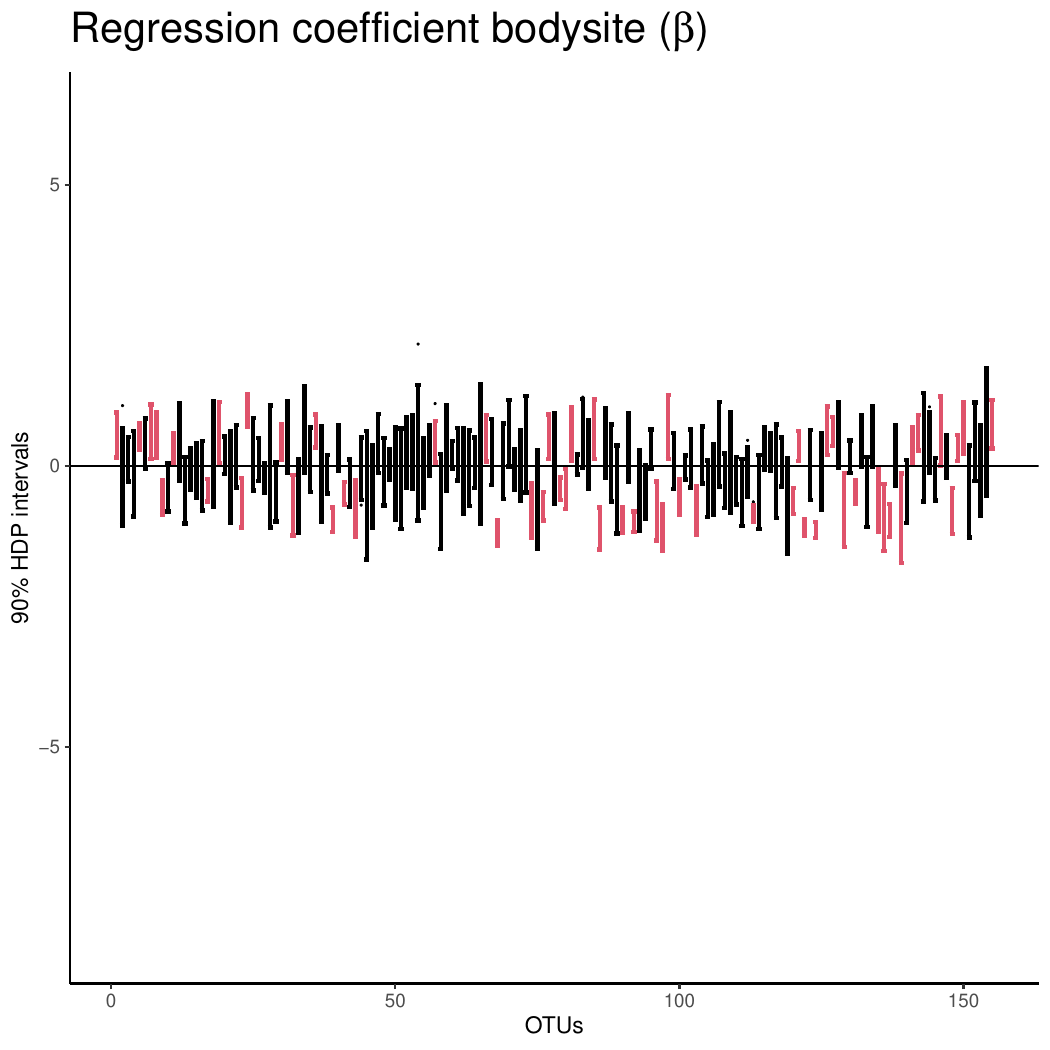}
\end{center}
\caption{90\% High Posterior Density (HPD) intervals of the $\theta$ (left) and $\gamma$ (right) regression parameter corresponding to the body site covariate, sorted according to the median of the $\theta$ regression coefficient across the posterior samples. Intervals that do not contain the zero are coloured in red.\label{fig:diffexp}}
\end{figure}

Including covariates in the inference of micribiota systems has the advantage that analyses that are typically conducted on a microbe by microbe basis, such as \cite{lee20}, are now naturally embedded in the overall joint model. Indeed, one can inspect the estimation and inference of any marginal effect of interest.  
In this particular analysis, there is interest in detecting the OTUs that are differentially expressed between the two different body sites. Figure \ref{fig:diffexp} shows how all 155 OTUs differ significantly between the two body sites.
Furthermore, the plots show how the regression coefficient of the $q$ parameter is highly significant, suggesting large differences between the proportion of zeros in the two environments for most OTUs.  In contrast to this, the regression coefficient of the $\beta$ parameter is less significant. This may in fact indicate that a simpler DW regression model, with a constant $\beta$ parameter, may be sufficient for some of the OTUs. As shown in the simulation (Figure \ref{fig:sim2}), these marginal effects, if left unaccounted for, could significantly distort the inference of the micriobiota system, which we discuss in the next section.

\subsection{Bayesian structure learning of the microbiota system}

We now turn to the main task of recovering the underlying network of dependencies between the OTUs. The space of possible graphs among 155 nodes is huge, creating a statistical and computational challenge at a level that has not been considered before in the context of Bayesian structure learning. Thus a few checks and considerations were made. Firstly, we start the MCMC chain by setting the initial graph to the empty graph, as we expect a sparse graph. Secondly, we perform the structure learning for a long number of iterations, namely 10 million MCMC iterations. Thirdly, we check the trace of the posterior edge probabilities and graph sizes for convergence. We also check the sensitivity to the graph prior, by setting the edge probability $\pi$ once to 0.2, the default value in \texttt{BDgraph}, and a second time to 0.04, which is the sparsity level of the graph detected by \texttt{rMAGMA} using zero-inflated negative Binomial as marginals and the stability selection criterion \texttt{stars} for model selection \citep{liu10}. Overall, we observe a high correlation among the edge posterior probabilities from the two chains (0.97). For the rest of the analysis, we consider the chain with $\pi=0.04$, which resulted in the highest log-likelihood when evaluated at the posterior estimates of the marginal distributions and precision matrix.

\begin{figure}[!h]
\begin{center}
\includegraphics[scale=0.4]{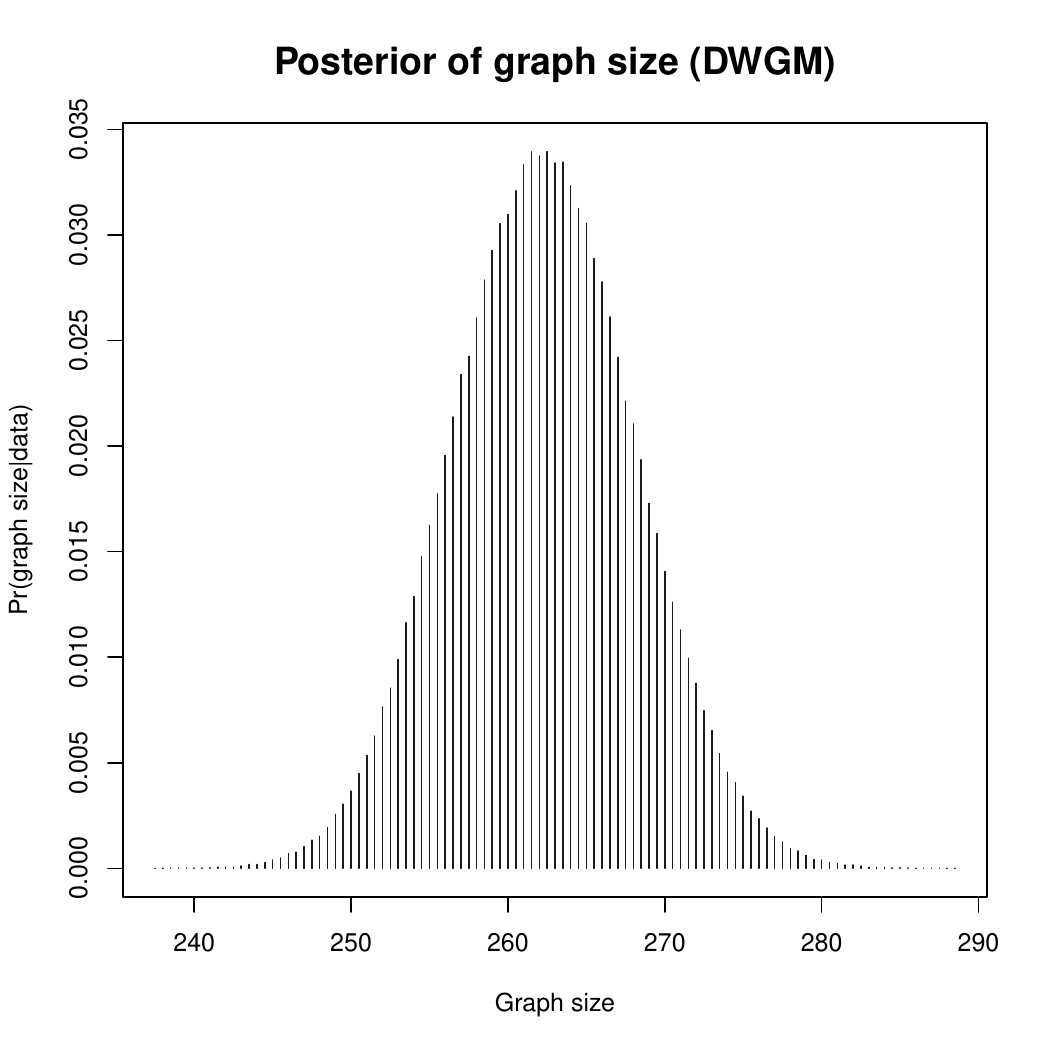}
\includegraphics[scale=0.4]{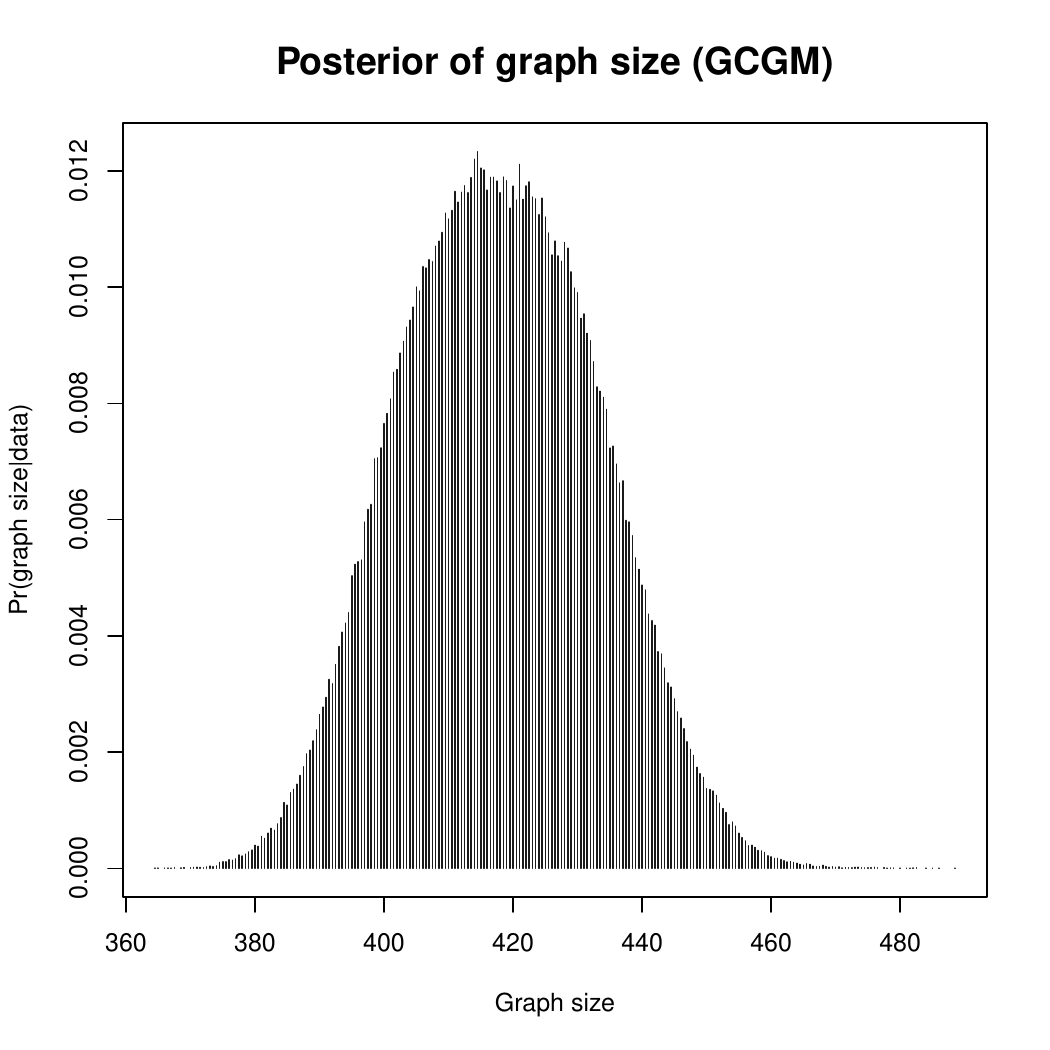}
\end{center}
\caption{Posterior distribution of graph sizes  for the model \texttt{DWGM} that accounts for covariates (left) versus the Gaussian copula graphical model that does not make use of covariates (\texttt{GCGM}, right).  \label{fig:graphsizes}}
\end{figure}

We firstly investigate the impact that the inclusion of covariates has on the inference of the underlying graph.  To this end, we compare our proposed approach with a Gaussian copula graphical model that uses the empirical distribution for the marginals (\texttt{GCGM}), i.e., a model that does not make use of covariate information.
We run also \texttt{GCGM} using 10 million iterations for the structure learning part and using the same prior on the graph ($\pi=0.04$). Figure \ref{fig:graphsizes} shows how the posterior distribution on graph sizes from the \texttt{DWGM} model (left) is concentrated on a sparser graph compared with the posterior distribution from the \texttt{GCGM} model. This was found also in the simulation study (Figure \ref{fig:sim2}) and suggests that  spurious links may have been detected by this second analysis where covariates have been omitted. 

\begin{figure}[!h]
\begin{center}
\includegraphics[scale=0.4]{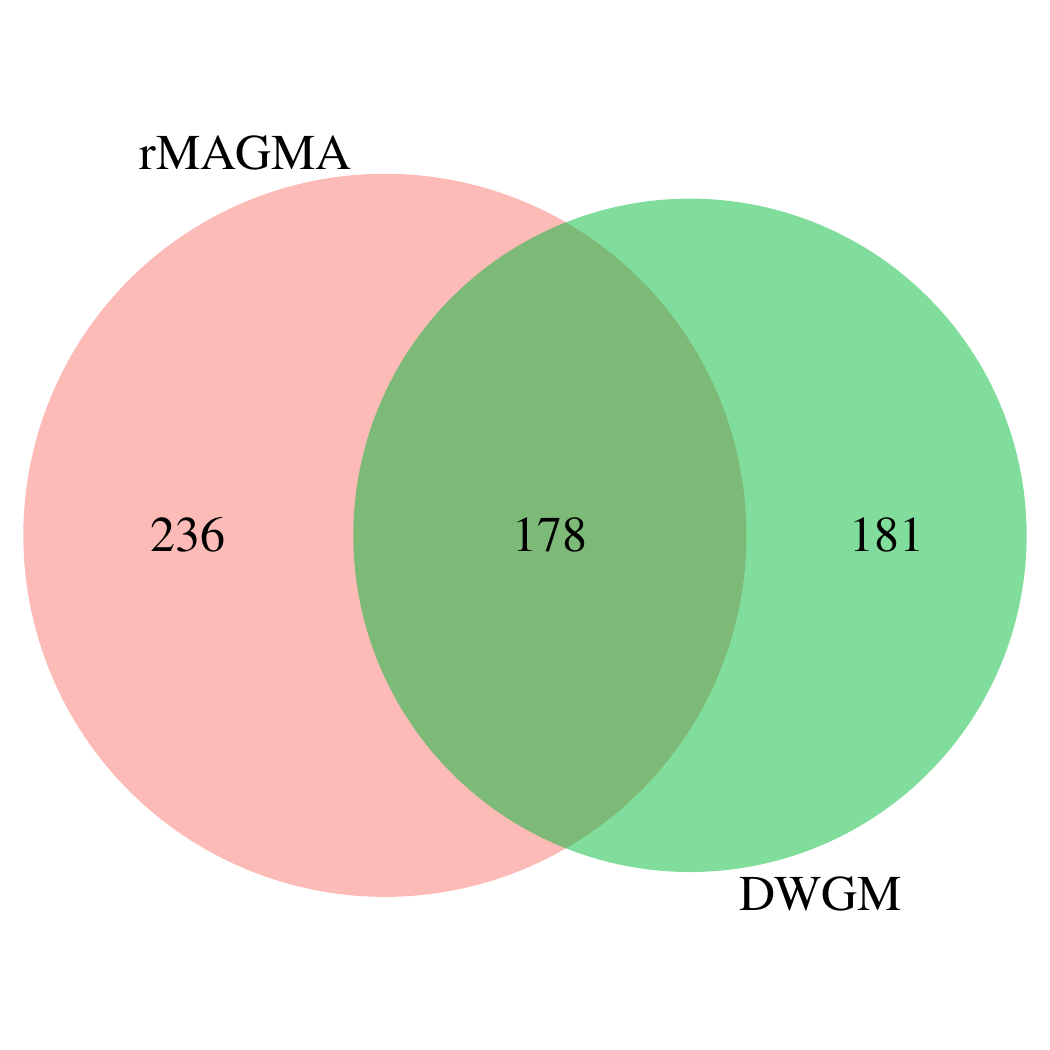}
\includegraphics[scale=0.4]{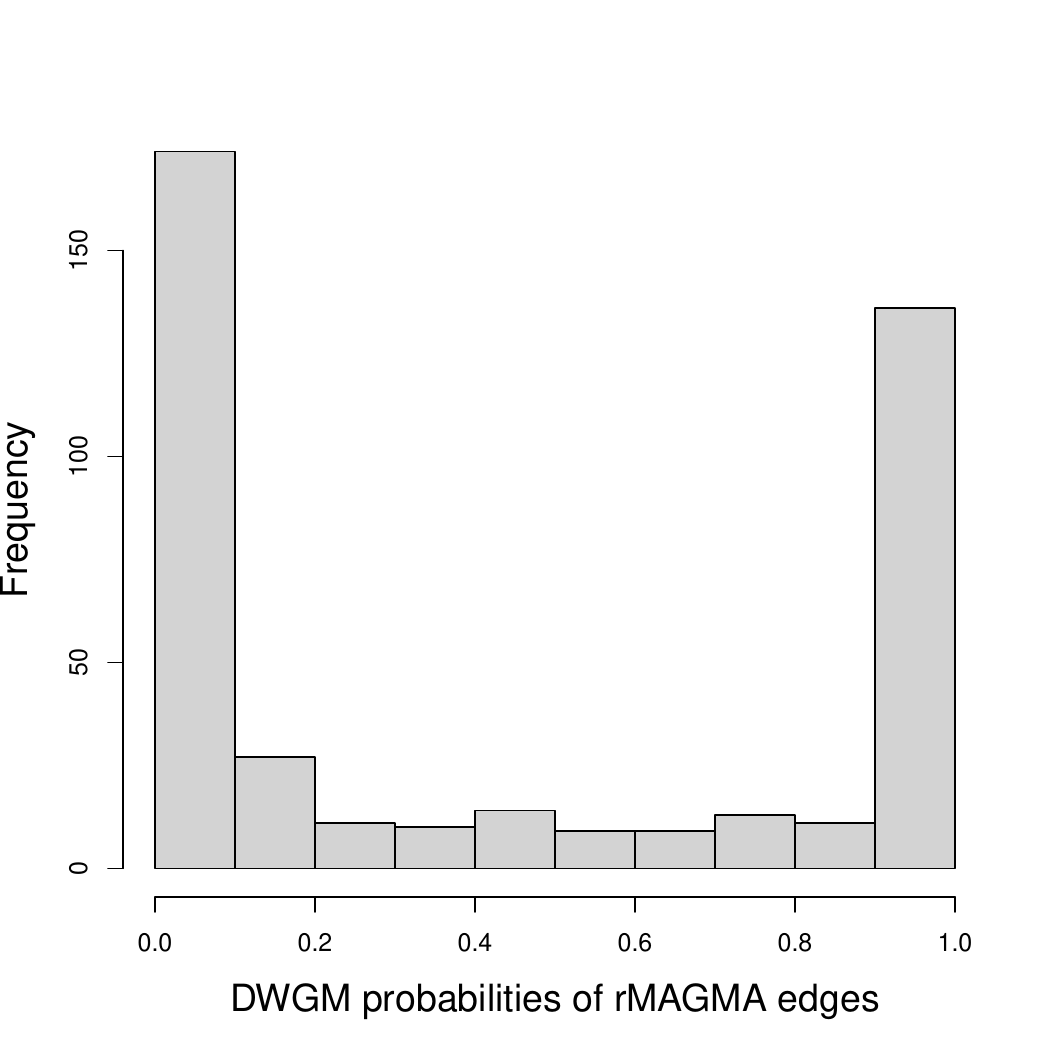}
\end{center}
\caption{Left: Venn diagram comparing the optimal graph estimated by \texttt{rMAGMA} using the \texttt{stars} criterion and the optimal graph estimated by \texttt{DWGM} by setting a cutoff of 0.5 on the posterior edge probabilities. Right: Histogram of \texttt{DWGM} posterior edge probabilities associated to the 414 edges detected by \texttt{rMAGMA}. \label{fig:venn}}
\end{figure}

Setting a cutoff of 0.5 on the edge posterior probabilities, the network contains 359 edges. Figure \ref{fig:venn} shows the overlap between these edges and the optimal graph detected by \texttt{rMAGMA} \citep{cougoul19}. 
There is a moderate (178 edges) overlap with the total of 414 edges detected by \texttt{rMAGMA}.
One of the major advantages of \texttt{DWGM}, which has not been considered before in the context of microbiome analyses, is that the uncertainty around the optimal graph is also measured. This is particularly important for structure learning in high dimensions, as noticed also in the simulation (Figure \ref{fig:sim3}). Indeed, the right plot of Figure \ref{fig:venn} shows how many of the edges detected by \texttt{rMAGMA} have a low posterior edge probability calculated by \texttt{DWGM}.

\begin{figure}[!h]
	\begin{center}
		\includegraphics[scale=0.5]{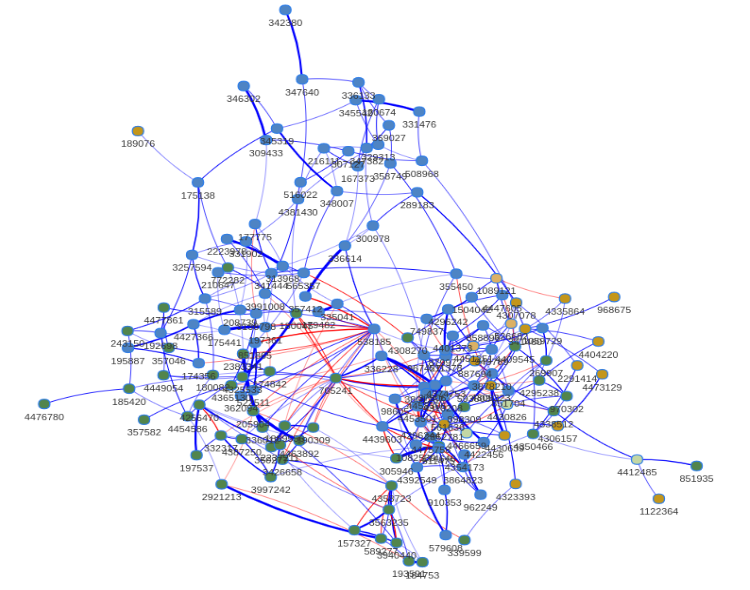}
	\end{center}
	\caption{The inferred mibrobiota system with posterior edge probabilities greater than 0.5. Node colours are associated to OTU phyla, with \tikzcircle[black, fill=NavyBlue]{2.5pt} Firmicutes, \tikzcircle[black, fill=PineGreen]{2.5pt}  Bacteobacteria, \tikzcircle[black, fill= Apricot]{2.5pt} Proteobacteria, \tikzcircle[black, fill=Goldenrod]{2.5pt} Actinobacteria and \tikzcircle[black, fill=YellowGreen]{2.5pt} Fusobacteria. Edge colors are associated to positive {\color{blue} \rule{0.5cm}{0.5mm}} and negative {\color{red} \rule{0.5cm}{0.5mm}} partial correlations, and width of the edges to their absolute values. \label{fig:dwgm_network} }
\end{figure}

Finally, Figure \ref{fig:dwgm_network} plots the network inferred by \texttt{DWGM}, with nodes coloured according to their phyla association (firmicutes, proteobacteria, bacteroidetes, actinobacteria, fusobacteria) and edges coloured according to their partial correlations, computed from the Bayesian averaging estimate of the precision matrix. 
The information on phyla association is useful in distinguishing between the stool and saliva microbiota. Indeed, firmicutes and bacteroidetes represent more than $90\%$ of the total human gut microbiota \citep{qin10} and have been found associated with several pathological conditions affecting the gastrointestinal tract, obesity and type 2 diabetes \citep{magne20,indiani18}. So we take this group of OTUs as representative of the gut microbiota. The remaining OTUs, with a mix of phylum levels, are instead associated to the saliva microbiota \citep{choi20}. In the optimal network (Figure \ref{fig:dwgm_network}), more connections are found within each group (average posterior edge probability equal to 4.7\% in gut and 10\% in saliva) than between the two groups (average posterior edge probability equal to 1.6\% between gut and saliva). While the connections are particularly strong within each group (average of posterior edge probabilities larger than 0.5  equal to 89\% in gut and 87\% in saliva, with associated average absolute partial correlation equal to 0.23 and 0.21, respectively), strong connections are detected also between the two groups (average of posterior edge probabilities larger than 0.5 between gut and saliva equal to 86\%, with associated average absolute partial correlations equal to 0.17), suggesting the presence of interactions between the two systems and supporting existing knowledge that oral microbes have the capacity to spread throughout the gastrointestinal system.

\section{Conclusion}
\label{sec:conclusion}
In this paper, we have presented a copula graphical modelling approach that is able to recover the core dependence structure from high dimensional and heterogeneous count data. We have shown the usefulness of this approach in learning interactions between microbes from count data provided by the latest micriobiome experiments, featuring high dimensionality, sparsity, heterogeneity and compositionality. The approach has three key features. 

Firstly, it allows to adjust for the effect of covariates in the marginal components of the model. This is useful, both in quantifying the effect of covariates of interest on the count variables and in aiding network recovery. The latter is down to two reasons: on one hand, the inclusion of covariates removes spurious dependencies that may be induced by the effect of the covariates on the variables of interest; on the other hand, the inclusion of (particularly continuous) covariates at the marginal level expands the region of support for consistent estimation of the copula in the case of discrete variables. 

Secondly, discrete Weibull regression is used for modelling the marginal distributions conditional on the covariates and is shown to be a simple (two parameters) yet flexible (broad dispersion levels) choice compared to more commonly used distributions for count data. Moreover, its definition as a discretized continuous Weibull distribution provides a latent continuous space in the vicinity of the data with a one-to-one mapping with the inferred conditional independence graph. This may be useful in deriving theoretical properties of the proposed approach. 

Thirdly, a Bayesian inferential procedure based on the extended rank likelihood and on an efficient continuous-time birth-death process allows to account for the full uncertainty both in the marginals, and thus in the covariate effects, and in the graph component. The latter is important, particularly in high dimensional settings where model selection methods for regularized approaches do not work well and where there is typically a large uncertainty around the optimal graph. The method proposed captures this uncertainty at the level of the graph
structure (via posterior probabilities of each link) and intensity of the interactions (via posterior estimates of partial correlations), but any other graph statistics of interest can be estimated via Bayesian averaging.

The simulation study and the real data analysis of microbiome data show the usefulness of the proposed approach at inferring networks from high-dimensional count data in general, and its relevance in the context of microbiota data analyses in particular. Indeed, the inferred interactions between firmicutes and bacteroidetes in the gut microbiota can create an opportunity for microbiome research to develop new microbial targets for the nutritional or therapeutic prevention and management of pathological conditions affecting the gastrointestinal tract, such as inflammatory bowel diseases, obesity and type 2 diabetes. At the same time,  the analysis proposed has shown the potential to detect crucial interactions between the gut and oral microbiota, which has been suggested only recently in the literature.

\section*{Software}
The method proposed in this paper is implemented in the R package \texttt{BDgraph} which is freely available 
from the Comprehensive R Archive Network (CRAN) at \url{http://cran.r-project.org/packages=BDgraph}.

\section*{Funding}
This project was partially supported by the European Cooperation in Science and Technology (COST) [COST Action CA15109 European Cooperation for Statistics of Network Data Science (COSTNET)].

\bibliographystyle{chicago}
\bibliography{references-DW}

\end{document}